\documentclass{aastex63}
\usepackage{bm}
\usepackage{amsmath}
\usepackage{color}
\usepackage{graphicx}
\received{June 1, 2019}
\revised{January 10, 2019}
\accepted{\today}
\submitjournal{ApJ}

\shorttitle{Malti-phase circumnuclear gases with low-$\beta$ B-field}
\shortauthors{Kudoh et al.}

\begin{document}

\title{Multiphase Circumnuclear Gas in a Low-$\beta$ Disk: Turbulence and Magnetic Field Reversals}

\correspondingauthor{Yuki Kudoh}
\email{k5751778@kadai.jp, yuki.kudoh@nao.ac.jp}

\author{Yuki Kudoh}
\affiliation{Graduate School of Science and Engineering, Kagoshima University, Korimoto, Kagoshima 890-0065, Japan}
\affiliation{National Astronomical Observatory of Japan, Mitaka, Tokyo 181-8588, Japan}

\author{Keiichi Wada}
\affiliation{Graduate School of Science and Engineering, Kagoshima University, Korimoto, Kagoshima 890-0065, Japan}

\author{Colin Norman}
\affiliation{Department of Physics and Astronomy, Johns Hopkins University, 3400 N. Charles Street, Baltimore, MD 21218, USA}
\affiliation{Space Telescope Science Institute, 3700 San Martin Drive, Baltimore, MD, 21218 USA}



\begin{abstract}

We studied the magnetic field structures and dynamics of magnetized multiphase gas on parsec scales
around supermassive black holes by using global 3D magnetohydrodynamics (MHD) simulations.
We considered the effect of radiative cooling and X-ray heating due to active galactic nuclei (AGNs).
The gas disk consists of a multiphase gas with (1) cold ($\leq 10^3$ K) and thin, and (2) warm ($\sim 10^4$ K) and thick components
 with a wide range of number densities.
The turbulent magnetic energy at maximum is comparable to
the thermal and turbulent kinetic energies in the turbulent motion.
We confirmed that the turbulent velocity of the warm gas in the ambient cold gas is caused by magnetoconvective instability.
The turbulent magnetic field due to magnetorotational instability (MRI) is developed in the disk, but
the mean toroidal magnetic field dominates and supports in a quasi-steady state, where
the plasma-$\beta$, the ratio between gas pressure and magnetic pressure, is low ($\beta < 1$).
As often seen in adiabatic MHD simulations of rotating disks,
the direction of the mean toroidal field periodically reverses with time even in multiphase gas structures.
The direction reversal is caused by magnetic flux vertically escaping from the
disk and by the combination of the MRI and the Parker instability.

\end{abstract}


\section{Introduction} \label{sec:1}

Magnetic field plays a crucial role in gas dynamics and multiphase gas structures in the central regions of galaxies where various dynamical structures, such as outflows, jets, and turbulent motions of interstellar medium are present on a wide dynamic range from the accretion disk to kiloparsec scales. It is suggested that the magnetic field in the central region of our Galaxy is
a few tens of $\mu$G to mG \citep[see, e.g.,][]{2009A&A...505.1183F, 2010Natur.463...65C, 2017ARA&A..55..111H}.
In particular, \citet{Hsieh2018} estimated the milligauss toroidal field and plasma-$\beta$, $\beta = 0.01 - 1$,
where $\beta$ is the ratio between thermal pressure and magnetic pressure,
\begin{eqnarray}
\beta \equiv \frac{P_{\rm g}}{P_{\rm B}} = \frac{P_{\rm g}}{|\bm{B}|^2/2}.
\end{eqnarray}
\citet{2010ApJ...722L..23N,2013ApJ...769L..28N}
found that the mean toroidal field extends to scale heights with the galactic latitude $|b|<0.4^{\circ}$ deduced from the Fe 6.7 keV line emission.

The circumnuclear magnetic field is also observed in some galaxies.
In the proto-typical type-2 Seyfert, NGC 1068,
the polarimetry in the infrared suggested that
the magnetic field is dominated by a toroidal field with a few ten mG and $\beta \sim 0.15$ \citep{2015MNRAS.452.1902L}.
Using the water vapor masers in NGC 4258 \citep{2005ApJ...626..104M}, the upper limit of the toroidal magnetic field $\sim 100$ mG is inferred.

Global magnetohydrodynamic (MHD) simulations of a rotating gas disk suggested that a turbulent field is developed due to magnetorotational instability \citep[MRI;][]{1991ApJ...376..214B}.
Using adiabatic MHD simulations for the Galactic center, \citet{2009PASJ...61..411M} and \citet{2015MNRAS.454.3049S} reported that magnetic turbulence is driven by MRI.
Turbulence contributes to the angular momentum and mass transport in the disk. In the nonlinear phase of MRI, it is characterized by a quasi-periodic reversal of the direction of the mean toroidal field
 \citep[e.g.,][]{Beckwith2011, 2011ApJ...736..107O, 2012ApJ...744..144F, 2013ApJ...764...81M, 2013ApJ...763...99P, 2016ApJ...826...40H}.
This field reversal is caused by the vertical transport of the mean field buoyantly from the midplane.
As a result, the amplification of the magnetic turbulence is saturated and, consequently, the field strength and
the plasma-$\beta$ in the mid-plane are limited in the nonlinear phase.

When $\beta \ga 1$, the time scale of the direction reversal is often observed to be about 10 rotational periods.
On the other hand, for $\beta < 1$,
this time scale is longer in the nonlinear regime depending on $\beta$.
Local 3D shearing box simulations reported that the reversal pattern is irregular
for the isothermal stratified gas disk \citep{2013ApJ...767...30B,2016MNRAS.457..857S}.
They also showed that when $\beta \la 0.4$, the field does not show reversal within 150 rotational periods.
Using global MHD simulations, \citet{2018ApJ...857...34Z} and \citet{2020MNRAS.492.1855M} studied the MRI
in a low-$\beta$ disk. Contrary to local MHD simulations, the plasma-$\beta$ is larger than unity at the midplane, and no field reversal is observed within 50 rotational periods.
\citet{2017MNRAS.467.1838F} conducted simulations starting from a strong toroidal magnetic field ($\beta=0.1$),
and they found that magnetic field dissipates to $\beta \sim 10$ in a steady state after 10 rotational periods.
\citet{2016MNRAS.460.3488S} pointed out that the poloidal field is necessary to form a strongly magnetized disk.
\citet{Begelman2007} studied the low-$\beta$ disk by compiling
 the typical unstable condition for MRI, and showed that
even if there was no vertical magnetic field, the low-$\beta$ MRI is limited by the unstable condition $\beta \ga c_s/v_{\rm K}$,
where $v_{\rm K}$ and $c_s$ are the Keplerian and sound speeds, respectively.
Therefore, for a given $v_{\rm K}$, the plasma-$\beta$ driving MRI is determined by the disk temperature,
and it is expected that an MRI with a low $\beta$ can develop in a low-temperature gas disk.

In most of the previous global MHD simulations of the magnetic field in a rotating disk,
the gas was assumed to be adiabatic; therefore, the multiphase nature of the magnetized gas was not well studied.
However, the circumnuclear gas on parsec scales should consist of cold ($T \lesssim 10^3$ K), warm ($T\sim 10^4$ K), and hot ($T \ga 10^5$ K) gases (see, e.g., for AGN: \citealt{2015ARA&A..53..365N}, and for the Galactic center: \citealt{2013ApJ...770...44L}).
Using global 3D simulations and considering radiative cooling and heating effects,
 \citet{Wada2009} and \citet{Wada2012c} showed that the gas inside tens of parsecs from SMBHs becomes multiphase.
They applied their model to the circumnuclear disk in the Circinus galaxy,
which is the nearest type-2 Seyfert galaxy, and found that the model is consistent with multi-wavelength observations
\citep{2016ApJ...828L..19W}, such as CO$(3-2)$ and [C{\footnotesize I}]$(1-0)$ emission lines \citep{2018ApJ...852...88W, 2018ApJ...867...48I}.
However, the magnetic field was not taken into account in their models.

There are several numerical studies for the magnetized circumnuclear gas on a pc scale.
\citet{Chan2017} and \citet{Dorodnitsyn2017} studied the evolution of warm gas with $\beta = 1$.
However, the timescale of their simulations is about 10 rotational periods, which is not long enough
for the magnetic field to become a steady state.
The long-term steady state behavior of the direction reversal of the mean magnetic field was not well studied.
Moreover, MRI with low $\beta$ in a nonlinear phase was not studied for the multi-phase gases, especially below $10^4$ K.
For example, it is not clear whether turbulence is maintained even in the cold gas,
and how different are the structures of the magnetic field compared to the adiabatic gas.
The structure of the circumnuclear gas with a magnetic field should be important to consider the effect of the radiation feedback
from the AGN.
In order to clarify these questions, we study the long-term behavior of the strong magnetized gas
around an SMBH by taking into account realistic cooling and heating processes.

This paper is organized as follows.
In \S \ref{sec:2}, we present the basic equation, initial condition, and numerical model.
Cooling and heating processes are described in \S \ref{subsec:21}.
Numerical results are shown in \S \ref{sec:3}.
Development of the MHD turbulence is show in \S \ref{subsec:31}.
Time evolution of the mean toroidal magnetic field and the physical origin of the turbulence are discussed in \S\S \ref{subsec:32} and \ref{subsec:33}. The thermal structures of the magnetized multiphase gas are presented in \S \ref{subsec:34}.
In \S \ref{sec:4}, we discuss the direction reversal with low $\beta$ (\S \ref{subsec:41}) and the radiation pressure (\S \ref{subsec:42}).
Finally, we summarize the results in \S \ref{sec:5}.

\section{Numerical Setup} \label{sec:2}

\subsection{Basic Equations} \label{subsec:21}

We study the pc-scale magnetized gas disk using 3D MHD simulations, considering radiative cooling and
various heating effects in the cylindrical coordinate $(R, \varphi, z)$.
The resistive MHD equations are:

\begin{eqnarray}
\displaystyle \frac{\partial  \bm{B} }{\partial t} + \bm{\nabla} \times \bm{E}=0, \label{eq:induction}
\end{eqnarray}
\begin{eqnarray}
\frac{\partial \rho}{\partial t} + \bm{\nabla} \cdot \left[\rho \bm{v} \right]  =0,
\end{eqnarray}
\begin{eqnarray}
 \displaystyle \frac{\partial}{\partial t} \left( \rho \bm{v} \right)
 + \bm{\nabla} \cdot \left[ \rho  \bm{vv} + \left( {P}_{\rm g}  + \frac{B^2}{8 \pi} \right) \bm{I} - \frac{\bm{BB}}{4 \pi} \right]
 =  - \rho \bm{\nabla} \Phi,     \label{eq:momentum}
\end{eqnarray}
\begin{eqnarray}
\displaystyle \frac{\partial}{\partial t} \left(e+ \frac{B^2}{8 \pi }  \right)
+ \bm{\nabla} \cdot \left[ \left( e + P_{\rm g} \right) \bm{v} -  \bm{E} \times \bm{B}  \right]
 = - \rho \bm{ v} \cdot \bm{\nabla} \Phi  - \rho {\cal L}(n, T),     \label{eq:energy}
\end{eqnarray}
\begin{eqnarray}
e \equiv \frac{P_{\rm g}}{\gamma_{{\rm g}} -1}  +\frac{1}{2} \rho v^2,
\end{eqnarray}
where $\bm{B}$ is the magnetic field, $\rho$ is the gas density, $\rho \equiv m_{\rm H} n\sim m_{\rm H} (n_n+n_i)$ with
number densities of neutral and ionized hydrogen. $P_{\rm g}$ is the thermal gas pressure.
The gas temperature is adopted as that of an ideal gas with the specific heat ratio $\gamma_{\rm g}=5/3$.
The gravitational potential is assumed to be the Newtonian potential, $\Phi = GM_{\rm BH}/r$, where $G$ is the gravitational constant, $M_{\rm BH}=10^7 M_{\sun}$ is the mass of the SMBH,
and $r=\sqrt{R^2+z^2}$ is the distance from the SMBH.
The electric field $\bm{E}$ obeys the Ohm's law, $\bm{E}= -\bm{v} \times \bm{B} + \eta \bm{\nabla} \times \bm{B} $.
We assume the anomalous resistivity $\eta$ as modeled by \citet{Yokoyama1994},
\begin{equation}
\eta=
\begin{cases}
\min \left\{\eta_{\rm max}, \eta_0 \left( v_d / v_c -1 \right)^2  \right\},  & v_d \geq v_c \\
0,                                               & v_d < v_c,
\end{cases}
\end{equation}
where $v_d= |\bm{\nabla} \times \bm{B}|/\rho$ is the electron-ion drift velocity, $v_c= 10^8$ cm s$^{-1}$ is the critical velocity, and we adopted $\eta_0= 10^{-9}$ and $\eta_{\rm max}=10^{-6}$ pc$^2$ yr$^{-1}$, respectively.
Note that the results presented below are not sensitive to $\eta_0$ and $\eta_{\rm max}$
(see also, the Appendix \ref{appendix}).

\begin{figure}[h!]
\epsscale{0.6}
\plotone{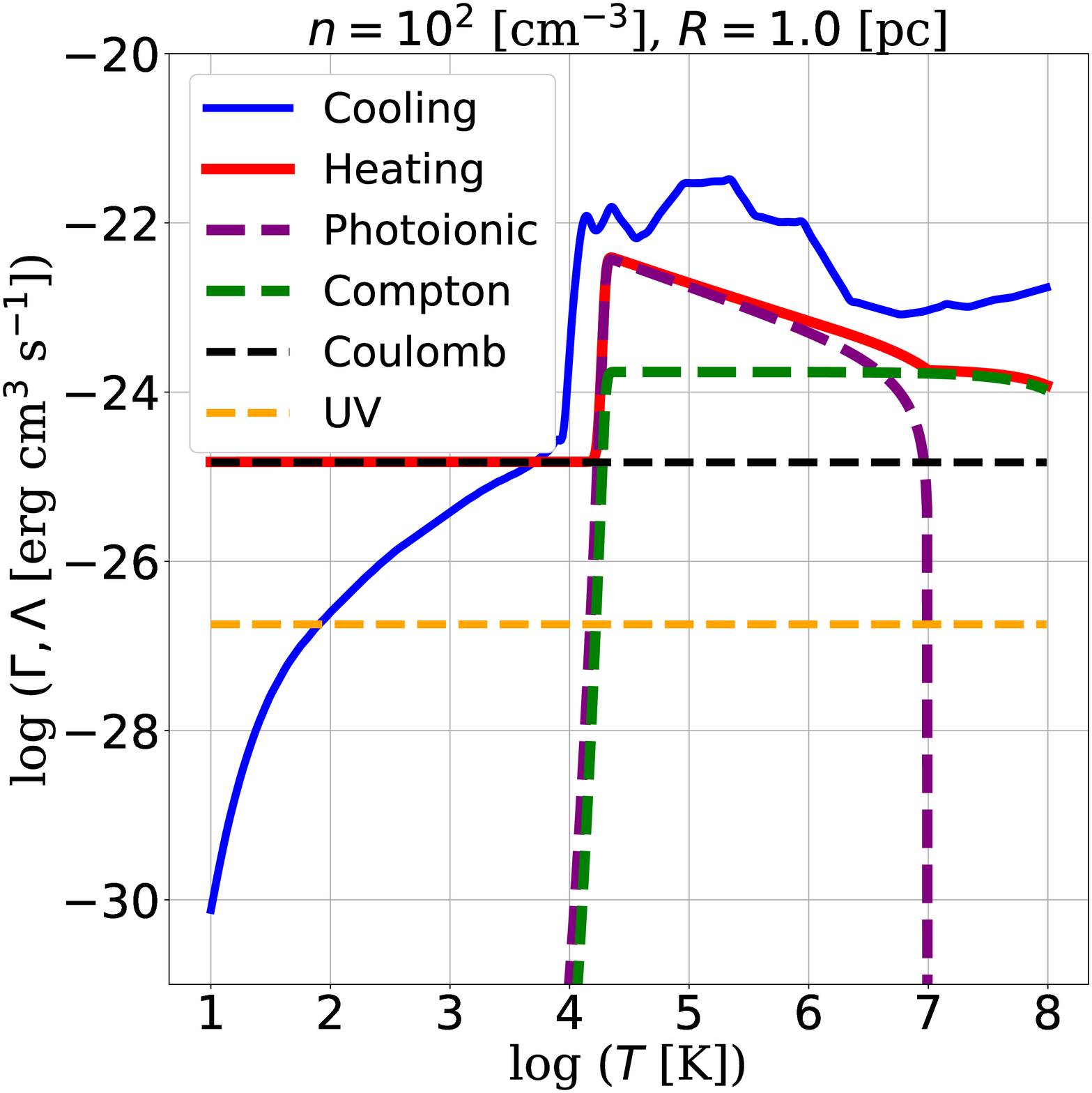}
\caption{
The cooling (blue) and the heating (red) functions for $n=10^2$ cm$^{-3}$.
The total heating rate is decomposed with various sources:
 photoionization (purple), Compton (green), Coulomb collision (black), and UV radiation (yellow).
\label{fig:Cool}}
\end{figure}

The radiative cooling and heating term $\rho {\cal L} $ in Equation \ref{eq:energy} is given as
\begin{eqnarray}
\rho {\cal L} = n^2 \Lambda -  n \left(\Gamma_{\rm UV} +\Gamma_{\rm X}  \right),
\end{eqnarray}
where the cooling function $\Lambda$ (Figure \ref{fig:Cool}) is taken from \cite{Meijerink2005} and \cite{Wada2009}.
As a major heating source, we consider that X-ray photons come from the accretion disk, and the heating function is $\Gamma_{\rm X} = \Gamma_{\rm Coulomb}+\Gamma_{\rm Compton}+\Gamma_{\rm photoionic}$.
The Coulomb interaction is given by
\begin{eqnarray}
\Gamma_{\rm Coulomb } \equiv  \eta_{\rm h} H_{\rm X}    ~{\rm erg~} {\rm s}^{-1},
\end{eqnarray}
where $\eta_{\rm h}$ is the efficiency in fixed $0.2$, and $H_{\rm X}$ is the X-ray energy deposition rate, $H_{\rm X}=3.8 \times 10^{-25} \xi {\rm ~erg~s}^{-1}$.
For the Compton and photoionization interactions, we use a formula given by \cite{Blondin1994}, i.e.
\begin{eqnarray}
\Gamma_{\rm Compton  } \equiv 8.9 \times 10^{-36} \xi n \left( T_X - 4 T \right)  ~{\rm erg~} {\rm s}^{-1},
\end{eqnarray}
\begin{eqnarray}
\Gamma_{\rm photoionic  } \equiv 1.5 \times 10^{-21} (\xi n)^{1/4} n^{3/4} T^{-1/2} \left(1 -  T/T_X \right)  ~{\rm erg~} {\rm s}^{-1},
\end{eqnarray}
where $T_{\rm X}=10^8$ K is the characteristic temperature of an X-ray photon.
Here the ionization parameter $\xi$ is
\begin{eqnarray}
\xi \sim 1.31 \times 10^2  \left( \frac{L_{\rm X}}{10^{-4} L_{\rm Edd} } \right) \left( \frac{r}{1 ~{\rm pc} } \right)^{-2}  \left( \frac{n}{ 10^2 ~{\rm cm}^{-3}} \right)^{-1} {\rm ~erg~cm~s}^{-1}, \label{eq:xi}
\end{eqnarray}
where $L_{\rm X}$ is X-ray luminosity and $L_{\rm Edd}$ is the Eddington luminosity for $M_{\rm BH}=10^7 M_{\sun}$.
In this simulation, we set $\xi=100$ for simplicity.

The photoelectric heating, assuming spatially uniform FUV is taken into account,
\begin{eqnarray}
\Gamma_{\rm UV  } = 1.8 \times 10^{-25}   {\rm erg~} {\rm s}^{-1}.
\end{eqnarray}

\subsection{Initial Condition and Normalized Unit}
We start the simulation from an equilibrium torus.
In order to construct the equilibrium torus solutions with a toroidal field, we assume the polytropic relation,
\begin{eqnarray}
P_{\rm g} = K \rho^{\gamma_{\rm g}}, \label{eq:polytropic}
\end{eqnarray}
where $K$ is the polytropic constant, and the weak toroidal field $B_{\varphi}$ is given by plasma-$\beta$, $\beta=100$.
To give the density distribution, we assume the rotation velocity to be the radial distribution of angular momentum $L(R)$,
\begin{eqnarray}
L(R)=L_0 \left(\frac{R}{R_0} \right)^a,   \label{eq:angular}
\end{eqnarray}
where $L_0(\equiv R_0 v_0)$ and $R_0$ adopt the normalized units and $a=0.35$ is the power index.
The flux surface $\Psi(R, z)$ is then given as
\begin{eqnarray}
\Psi(R, z) \equiv \Phi(R, z) + \frac{L^2(R)}{2(1-a) R^2} + \frac{ 1 }{\gamma_{\rm g} -1 } c_s^2 (R, z) + \frac{\gamma_{\rm g}  }{2 \left( \gamma_{\rm g} -1 \right) } v_A^2 (R, z),
\end{eqnarray}
where $c_s^2 = \gamma_{\rm g} P_{\rm g} / \rho $ is the square of the sound speed
and $v_{A}^2= B_{\varphi}^2/(4\pi \rho)$ is the square of the Alfv\'{e}n speed.
We defined the normalized unit as the torus density maximum, $\Psi_0 = \Psi (R_0, 0)$.
For the condition of positive pressure, we can obtain a torus density distribution $\Psi(R,z)= \Psi_0$,
\begin{eqnarray}
\rho_{\rm torus}(R, z)=\rho_{0} \left[ \frac
{\max \left\{ \Psi_0 - \Phi(R, z) - \frac{L(R)^2}{2(1-a)R^2}, 0  \right\}}
{K \frac{\gamma_{\rm g}}{\gamma_{\rm g} -1} \left( 1 +  \beta^{-1} R^{2(\gamma_{\rm g} -1)} \right) }
\right]^{1/(\gamma_{g} - 1)}.
\end{eqnarray}
We assume that the torus is embedded in the isothermal nonrotating and nonmagnetized halo, with pressure and
density given by
\begin{eqnarray}
P_{\rm g, h} &=& K_{\rm h} \rho_{\rm h}, \\
\rho_{\rm h} &=& 10^{-5} \rho_0 \exp \left[ - \left\{\Phi(R, z) - \Phi(R_0,0) \right\}/K_{\rm h} \right].
\end{eqnarray}

The normalized quantities are listed in Table \ref{tab:nml}.
The polytropic indexes $K$ and $K_{\rm h}$ are parameterized by the square of the velocity ratio, $\varepsilon= K \rho_{0}^{\gamma_{\rm g} -1} / v_0^2 =\gamma_{\rm g}^{-1} c_{s}^2 (R_0,0)/  v_0^2  $ in the normalization, and we set $\varepsilon=0.04$ in the torus and $\varepsilon_{\rm h}=2$ in the halo, respectively.

\begin{deluxetable*}{llll}[h!]
\tablecaption{Normalization \label{tab:nml}}
\tablecolumns{4}
\tablenum{1}
\tablewidth{0pt}
\tablehead{
\colhead{Quantity} & \colhead{Unit} &
\colhead{Definition} & \colhead{Normalization}
}
\startdata
Central BH Mass &  $M_{\odot}$  & $M_{\rm BH}$ & $10^7 M_{\odot}$  \\
Length  & $R_0$ & $-$  & 1 pc  \\
Velocity & $v_0$ & $\sqrt{G M_{\rm BH} / R_0}$  & $2.07 \times 10^7 $  cm s$^{-1}$  \\
Time     & $t_0$ & $R_0/v_0$  & $4.74 \times 10^{3}$ yr  \\
Density & $\rho_0$ & $-$  & $1.673 \times 10^{-22}$ g cm$^{-3}$   \\
Pressure & $P_{\rm g0}$ & $\rho_0 v_0^2$  & $7.16 \times 10^{-8}$ erg cm$^{-3}$  \\
Temperature & $T_0$ & $v_0^2 ~ m_p/k_B$  & $5.19 \times 10^{6}$ K  \\
Magnetic Field & $B_0$ & $\sqrt{4 \pi \rho_0 v_0^2}$  & 949 $\mu$G  \\
\enddata
\end{deluxetable*}

\subsection{Numerical Methods and Model Setup} \label{subsec:23}
We use \texttt{CANS+} code \citep{2019PASJ...71...83M}, which is implemented using the HLLD solver \citep[Harten-Lax-van Leer discontinuitues;][]{Miyoshi2005} with div $B$ cleaning \citep{Dedner2002} and three-stage total variation diminishing Runge-Kutta time integration (TVDRK).
The fifth-order accuracy in space is achieved through the Monotonicity Preserving method \citep[MP5;][]{Suresh1997} to capture small-scale magnetic fluctuations.
Basic equations are solved using conservative forms with geometrical and gravitational source terms explicitly included.
After TVDRK updating, the cooling and heating source terms are treated in the implicit operator splitting approach.

As a numerical constraint in our simulation, we set the floor of the gas pressure in the grid cells where the minimum temperature is $T=20$ K or the minimum plasma-$\beta$ is $\beta=0.001$.
Comparing the volume-averaged energies of the grid cells where the lower limit is applied, the thermal energy is always 2-3 orders of magnitude smaller than the kinetic and magnetic energies.
This implies that the artificial thermal energy due to the numerical floor does not affect the gas dynamics.

The size of the simulation box is $0 \leq R/R_0< 12$,  $0<\varphi$ [rad]$< 2\pi$, $|z/R_0|<3$.
The grid size is $\Delta R = 0.01 R_0$ and $\Delta z =0.002 R_0$.
We use a coarser grid size outside for $R/R_0 > 2.0$ or $|z/R_0| > 0.4$, and for the region around the axis $R/R_0 < 0.2$.
The numbers of grid cells are $N_R = 256$, $N_{\varphi} = 512$, and $N_{z}=512$.

We assume the outflow boundary condition for the outer boundaries (i.e. $R/R_0=10$ and $|z|/R_0>2.8$).
For the azimuthal direction, the periodic boundary condition is assumed.
Meshes around the cylindrical axis are sent to the opposite computational domain of the azimuthal direction, which means that fluid can flow across the polar axis ($R=0$).
In the central region for $r/R_0 < 0.4$, we impose an absorbing boundary condition, i.e.
\begin{eqnarray}
q_{\rm new} =q - D(q-q_{\rm init}),
\end{eqnarray}
where $q$ and $q_{\rm init}$ are the primitive variables of the TVDRK updating state and the initial state, respectively.
Damping function $D(r)$ is modeled as,
\begin{eqnarray}
D(r)= 0.1 \left[ 1- \tanh \left(\frac{r- 0.2 R_0}{0.01 R_0} \right) \right].
\end{eqnarray}

We start the simulations assuming the adiabatic MHD, then
after the magnetic field strength sufficiently develops and the system becomes
a quasi-steady state at $t \leq 0.477$ Myr (25 rotational periods at $R=1$ pc),
the cooling and heating terms are considered until $t \leq 3.64$ Myr (97 rotational periods).

\section{Results} \label{sec:3}
\subsection{Development of the MHD Turbulence in the Torus}  \label{subsec:31}

\begin{figure*}[h!]
\begin{interactive}{animation}{fig2.mp4}
\epsscale{1.15}
\plotone{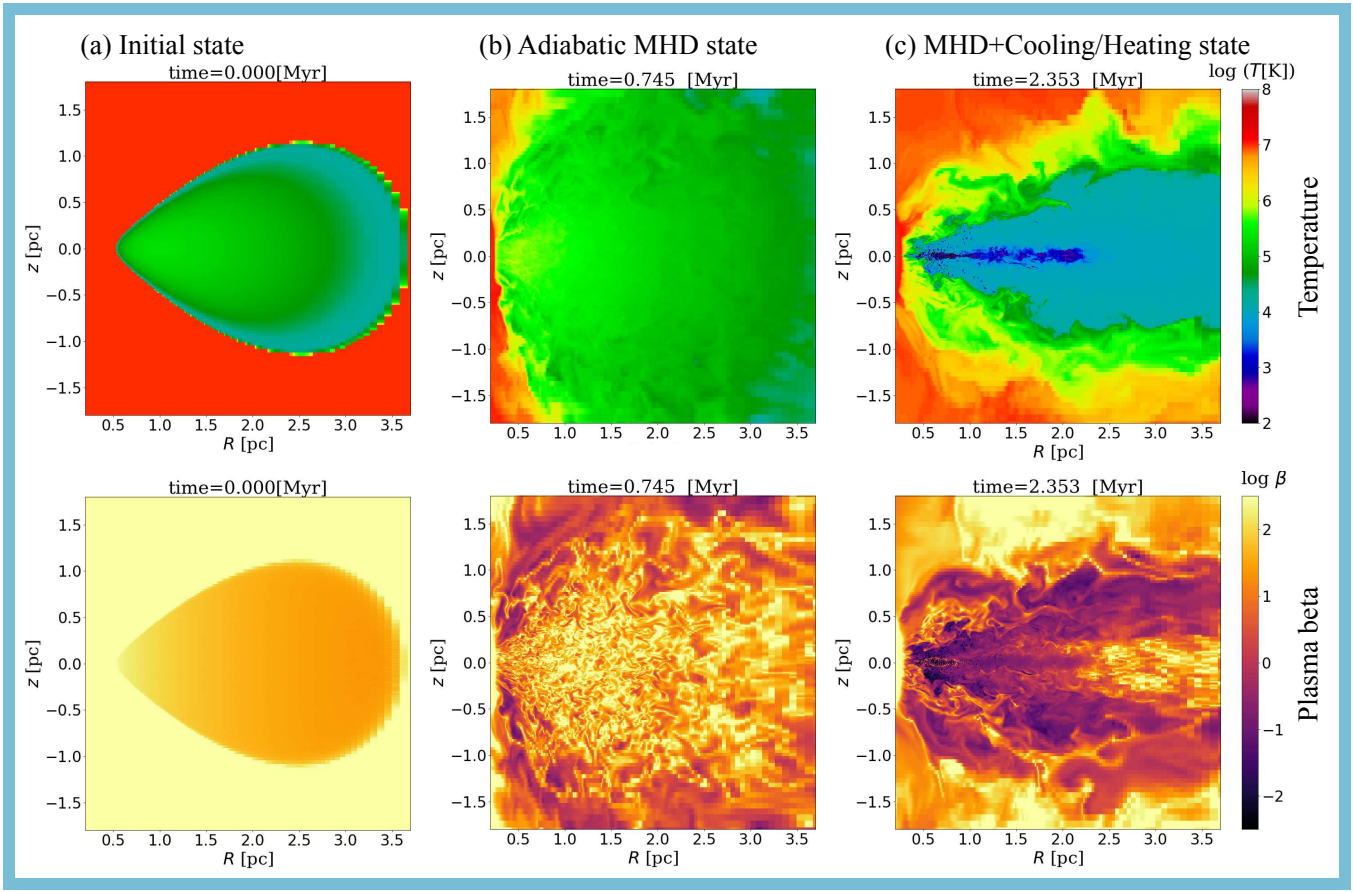}
\end{interactive}
\caption{Snapshots of temperature and plasma-$\beta$ in the $R$-$z$ plane.  Panels from left to right denote (a) initial state at $t=0.000$ Myr, (b) adiabatic MHD evolution at $t=0.744$ Myr, and (c) MHD including cooling and heating effects at $t=2.353$ Myr.  Associated with these snapshots we show an animation with temperature (top) and plasma-$\beta$ (bottom) at \url{https://astrophysics.jp/MHD_torus}.
 \label{fig:te}
 }
\end{figure*}

Figure \ref{fig:te} shows three snapshots ($t = 0.000, 0.744$ and 2.353 Myr) of the gas temperature and the plasma-$\beta$ on
a $R$-$z$ plane. Figure \ref{fig:te}a shows the initial conditions. As explained in Section \ref{subsec:23},
the system evolves adiabatically until $t = 0.477 $ Myr.
During this period, gas spreads out vertically, and intense magnetic field fluctuations are developed by the MRI and
 $\beta >0.6$ inside the torus.
 The magnetic field is stronger near the surface of the torus
with $\beta \sim 0.2$.
MRI causes not only turbulent motion but also heating due to magnetic reconnection.
After cooling and heating are taken into account,
the structure of the torus changes (Figure \ref{fig:te}c).
The plasma-$\beta$ around the midplane becomes smaller with $0.003<\beta< 6.0$.
The torus becomes geometrically thinner, and it consists of two components: cold disk ($T \la 10^3$ K) and warm disk ($T = 10^{3-5}$ K).
The cold, thin disk ($R < 2$ pc) is supported vertically by the magnetic field as discussed below.

\begin{figure*}[h!]
\epsscale{1.15}
\plotone{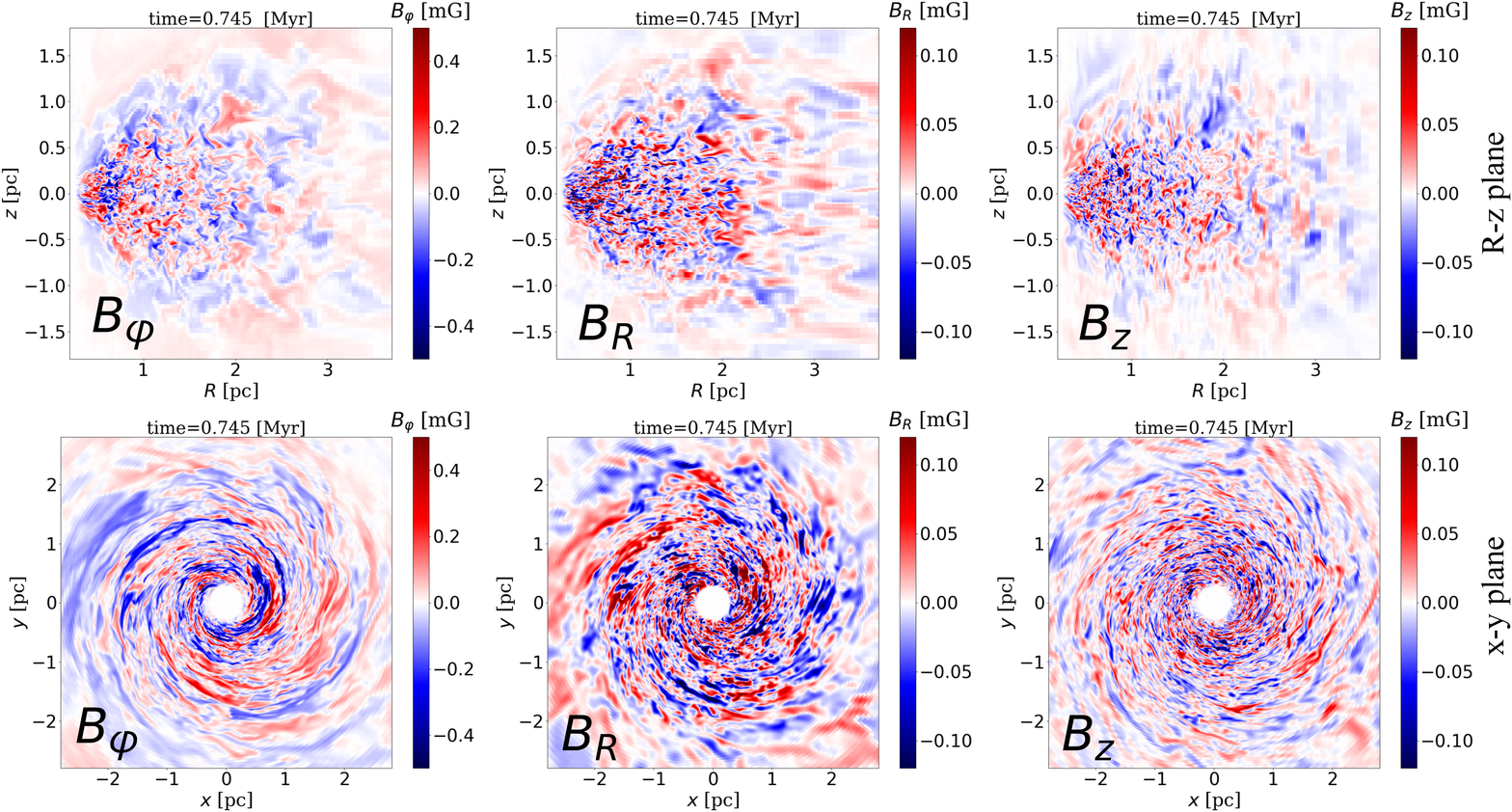}
\caption{Adiabatic MHD state of magnetic field distributions in the $x-y$ plane of $z=0$ (top) and the $R$-$z$ plane of $\varphi=0$ (bottom) at $t=0.744$ Myr.
From left to right, field components are toroidal, radial, and vertical fields.
Blue and red colors denote the negative and positive signs, respectively, for these right-handed coordinates.
\label{fig:mag_0250}}
\end{figure*}

Figure \ref{fig:mag_0250} shows the magnetic field structure at $t=0.744$ Myr of the adiabatic MHD state.
The turbulent field in the torus dominates in the $R$-$z$ and $R$-$\varphi$ planes.
The toroidal field ($B_{\varphi}$) shows the flux bundle of the positive direction (red) around the torus surface at $|z|=1.5$ pc.

\begin{figure*}[h!]
\begin{interactive}{animation}{fig4.mp4}
\epsscale{1.15}
\plotone{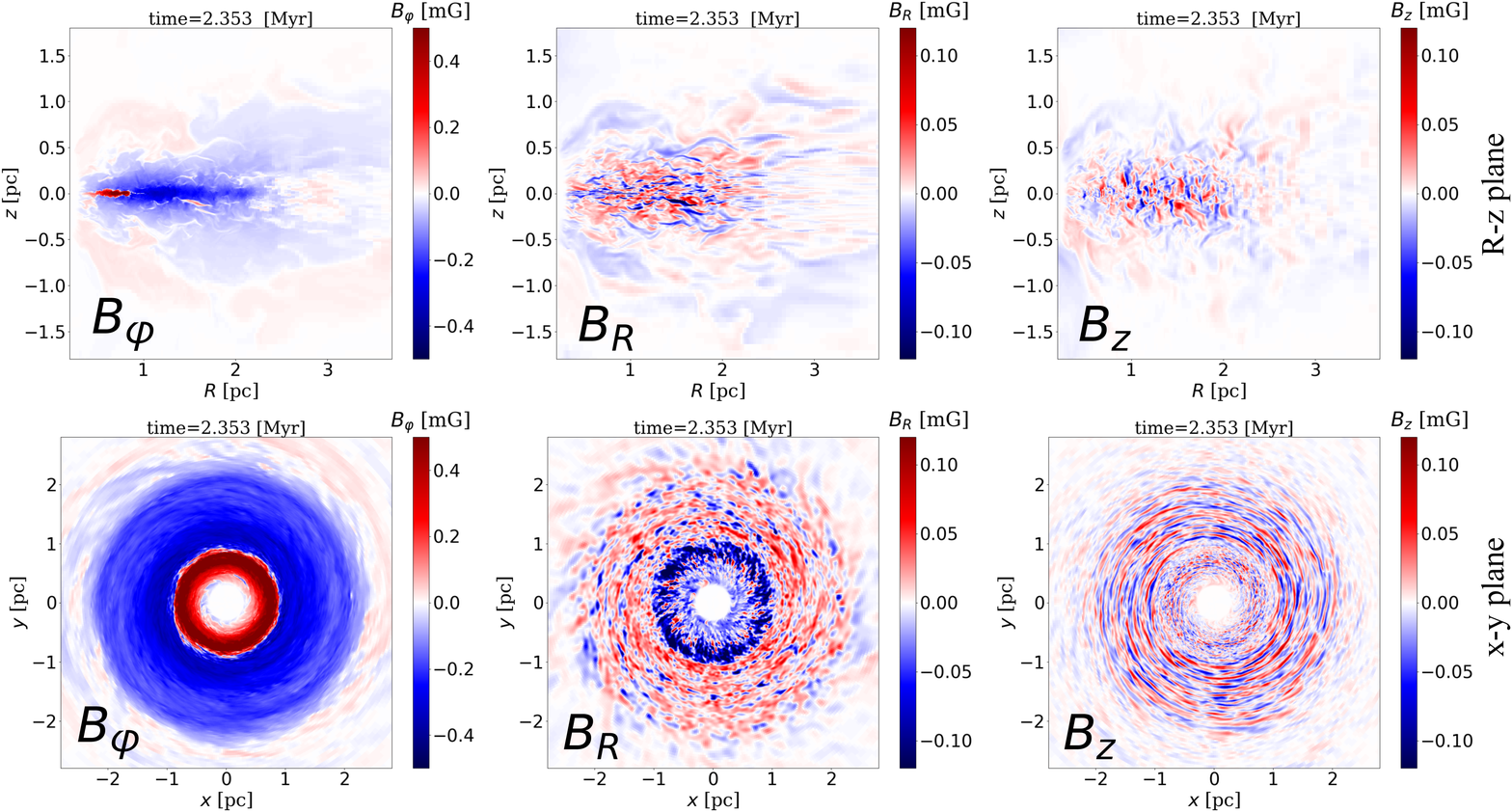}
\end{interactive}
\caption{
The same of snapshots in Figure \ref{fig:mag_0250}, but for the calculation of MHD with cooling/heating effects at $t=2.353$ Myr.
This figure is available as an animation,
running from $t=0.744$ to $t=3.642$ Myr.
This figure is available as an animation at \url{https://astrophysics.jp/MHD_torus}.
\label{fig:mag_0790}
}
\end{figure*}

Figure \ref{fig:mag_0790} is the same as Figure \ref{fig:mag_0250}, but cooling and heating are considered ($t = 2.353 $ Myr).
The field strength and turbulent fluctuation are markedly different from those in Figure \ref{fig:mag_0250}.
The toroidal field $B_\varphi$ dominates the total magnetic field.
The two plots of $B_\varphi$ show that
the mean toroidal field has more coherent structures compared to the that in the adiabatic phase (Figure \ref{fig:mag_0250}) with
opposite directions shown in blue and red in Figure \ref{fig:mag_0790}.
On the other hand, radial and vertical magnetic fields are dominated by the turbulent component.
We can also see that the patches of $B_R$ and $B_z$ with opposite directions tend to extend radially and vertically, respectively.

\begin{figure*}[h!]
\epsscale{1.2}
\plotone{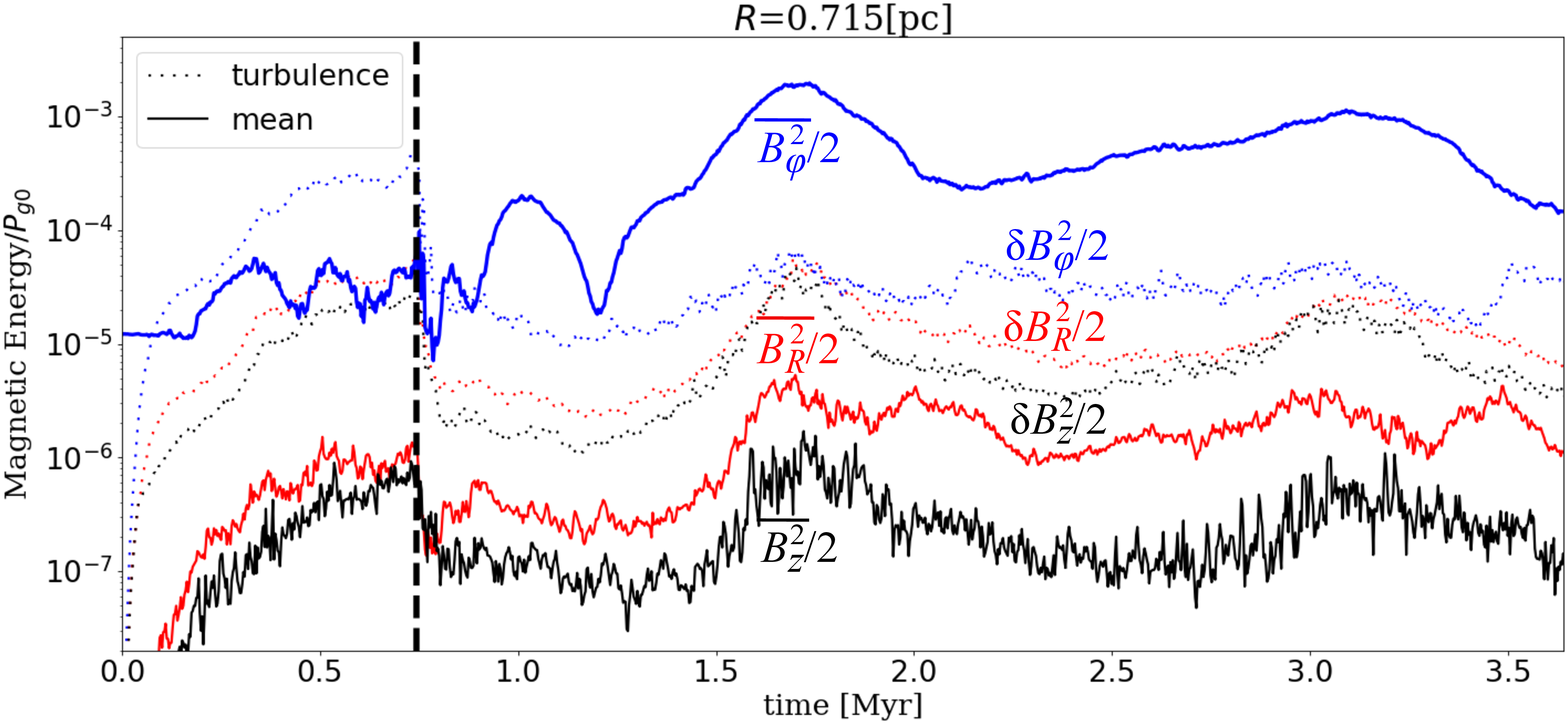}
\caption{
Time evolution of magnetic field energy separated into mean field ($\overline{B}_i^2/2$ where $i=R, \varphi, z$) and turbulent field ($\delta B_i^2/2$).
Each energy measures the volume weighted average over a ring of a rectangular cross section; $0.715-\Delta R/2 \le R {\rm ~pc} < 0.715+\Delta R/2$, $0 \le \varphi {\rm ~rad}<2 \pi$, and  $|z|  {\rm ~pc} \le 1$.
The dashed vertical line at $t= 0.744$ Myr is the turning point of the states between the adiabatic MHD (left side) and MHD+cooling/heating (right side).
Colors denote the toroidal component (blue), the radial component (red), and the azimuthal component (black).
\label{fig:512Emag}}
\end{figure*}

For a more quantitative observation, we decomposed the magnetic energy into mean and turbulent
components.
We measured the mean component as the azimuthal average,
\begin{eqnarray}
\overline{f} (R, z, t) \equiv \frac{1}{2 \pi} \int_{0}^{2 \pi} d \varphi f(R, \varphi, z, t).
\end{eqnarray}
Hence, the turbulent component is derived as,
\begin{eqnarray}
\delta f(R, \varphi, z, t) \equiv  f(R, \varphi, z, t) - \overline{f} (R, z, t),
\end{eqnarray}
where we notate the mean as $\bar{~}$ and the turbulence as $\delta$.
Figure \ref{fig:512Emag} shows the evolution of the magnetic energy in each component.
It is clear that the structures of the magnetic field change drastically after the cooling and heating.
For $t<0.5$ Myr, turbulent fields become exponentially stronger than mean fields, while the ratios between them are approximately constant,
i.e.  $\delta B_{\varphi}^2/\delta B_{R}^2  \sim 10$ and  $\delta B_{\varphi}^2/\delta B_{z}^2 \sim 22$.
After cooling/heating at $t=0.744$ Myr, the turbulent fields decrease quickly by one order of magnitude,
and they survive until $t \sim 3.5$ Myr. This turbulent field dissipation is caused by the decrease in the gas temperature (see in Section \ref{sec:4}).
During this phase, we found that $|\delta B_{\varphi}| > |\delta B_{R}| > |\delta B_{z}|$, but the ratios between them change.

In the adiabatic MHD state, the amplitude of $\overline{B}_{\varphi}$ (blue solid line in Figure \ref{fig:512Emag}) reaches a quasi-steady state with a periodical cycle of $T_{\rm cycle} = 0.163$ Myr $\sim 9.1$ rotational periods.
After cooling/heating effects are included, oscillating amplification of $\overline{B_{\varphi}}$ continues for $0.744<t<1.75$ Myr.
The quasi-steady state is achieved when $\overline{B_{\varphi}}$ does not significantly change .
However, $|\overline{B_{\varphi}}|$ shows quasi-periodic oscillations on a long time scale beyond reaching steady state.
This amplitude and period become larger and longer than those of the adiabatic MHD state.
Around the maximum of $|\overline{B_{\varphi}}|$ (i.e. $t=1.75, 3.08$ Myr), $|\delta \bm{B}|$ and $|\overline{B_{R}}|$ are maximized, and around the minimum of that (i.e. $t=2.1, 3.2$ Myr), $|\delta \bm{B}|$ and $|\overline{B_{R}}|$ are maximized.
We show the quasi-periodic spatial changing in the animation of Figure \ref{fig:mag_0790}.

Steady-state behavior is also seen from examining the mass flux inside 0.9 pc for $t>1.75$ Myr.
The radial profile of the net mass accretion rate appears to be roughly constant in time, which has been observed in simulation studies \citep[e.g.][]{Stone2001, Jiang2019}.
We took the time averages over 1.6 Myr, which is  the longest time scale in the periodic oscillation of $|\overline{B_{\varphi}}|$.
The time variation of the mass accretion rate due to MRI turbulence has been discussed in \cite{2011ApJ...736..107O}, \cite{2011ApJ...738...84H}, and  \cite{2016ApJ...826...40H}.
Since the mass flux varies with the averaging time interval, the time domain $0.744<t<1.75$ Myr is not steady.


\subsection{Direction Reversal and Vertical Transport} \label{subsec:32}

\begin{figure*}[h!]
\epsscale{1.2}
\plotone{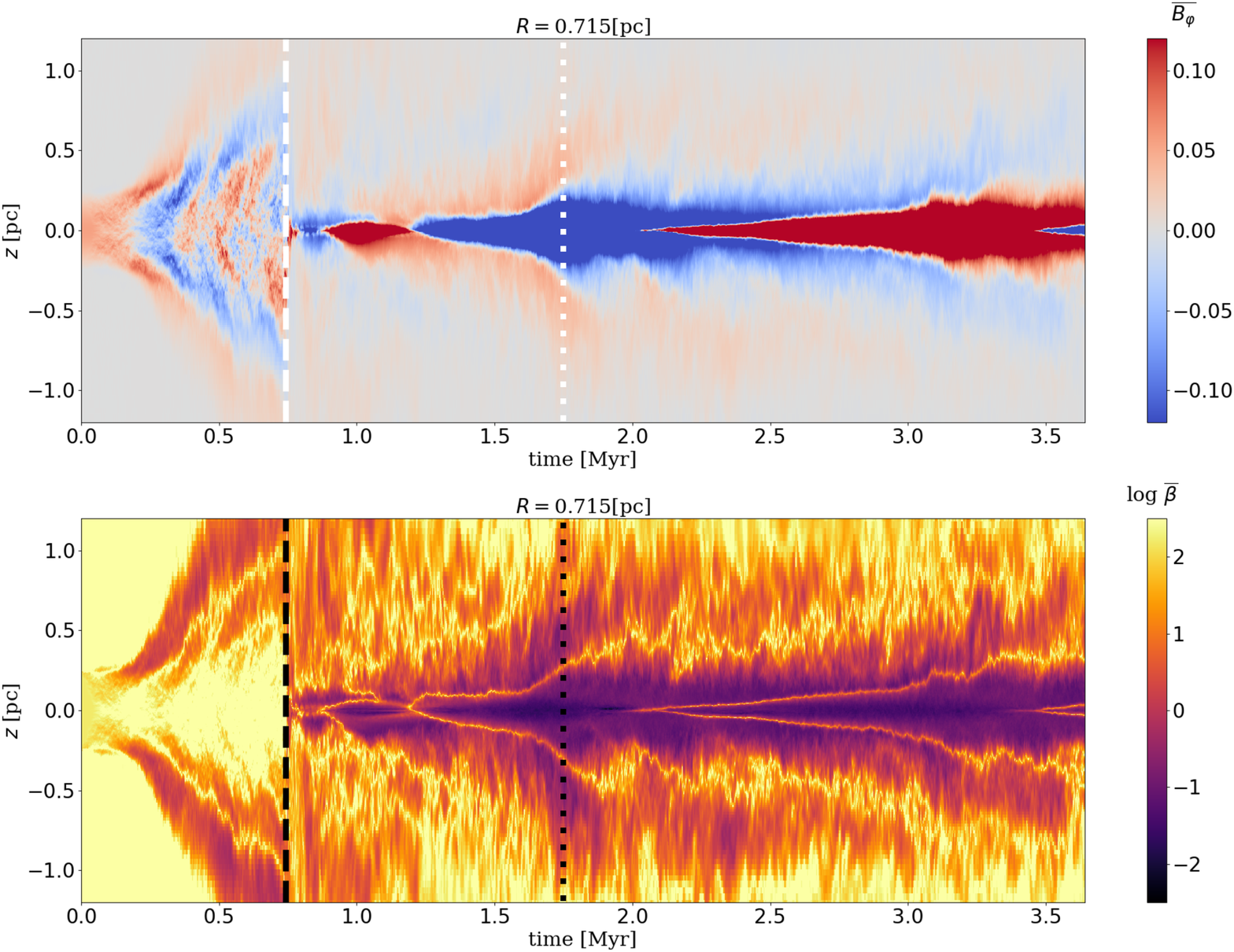}
\caption{
Space-time diagrams at $R=0.715$ pc.
Top panel is the mean of the toroidal field and the bottom panel is the mean plasma-$\beta$.
Dashed vertical line at $t= 0.744$ Myr is the turning point between the adiabatic MHD (left side) and MHD+cooling/heating (right side).
The Dotted vertical line in each panel at $t=1.75$ Myr is when the system becomes a quasi-steady state.
\label{fig:512bfm}}
\end{figure*}

The periodical cycle of the mean toroidal field energy is observed to have the pattern of a quasi-periodic direction reversal.
Figure \ref{fig:512bfm} (top), the so-called the butterfly diagram,
 is the space ($z$-direction) time evolution of the mean toroidal field direction at $R=0.715$ pc.
It shows that the direction periodically changes at a given $z$, and it also implies that the $B$-field escapes from the midplane.
Figure \ref{fig:512bfm} (bottom) is for the mean plasma-$\beta$, $\overline{\beta} \equiv 2 \overline{P_{\rm g}} / \overline{|\bm{B}|^2}$.
The disk surface traces the bounding surface for $\overline{\beta} \le 10$.
This implies that the magnetic field becomes amplified near the midplane and is transferred to the high latitudes of the disk.
The quasi-steady radial mass flow indicates the saturation of the MRI-driven turbulence.

In the MHD+cooling/heating state ($t > 0.744$ Myr), the lowest plasma-$\beta$ is in the midplane, and mean field transport is slower than
that of the adiabatic MHD. The direction of $B_{\varphi}$ in $1.20 < t$ Myr $< 2.27$ is vertically stratified in blue
around the mid-plane and red above that. The vertical transport changes from slow to fast at time $t = 1.74$ Myr.
Recall that the energy of $B_{\varphi}$ increases and decreases in Figure \ref{fig:512Emag}, and this time is the maximum point.
The same phenomenon occurs on the next reversal where it is red around the midplane and blue above that. We have confirmed the cycle through
the mean toroidal field direction reversal and escape from the disk.

\begin{figure*}[h!]
\epsscale{1.2}
\plotone{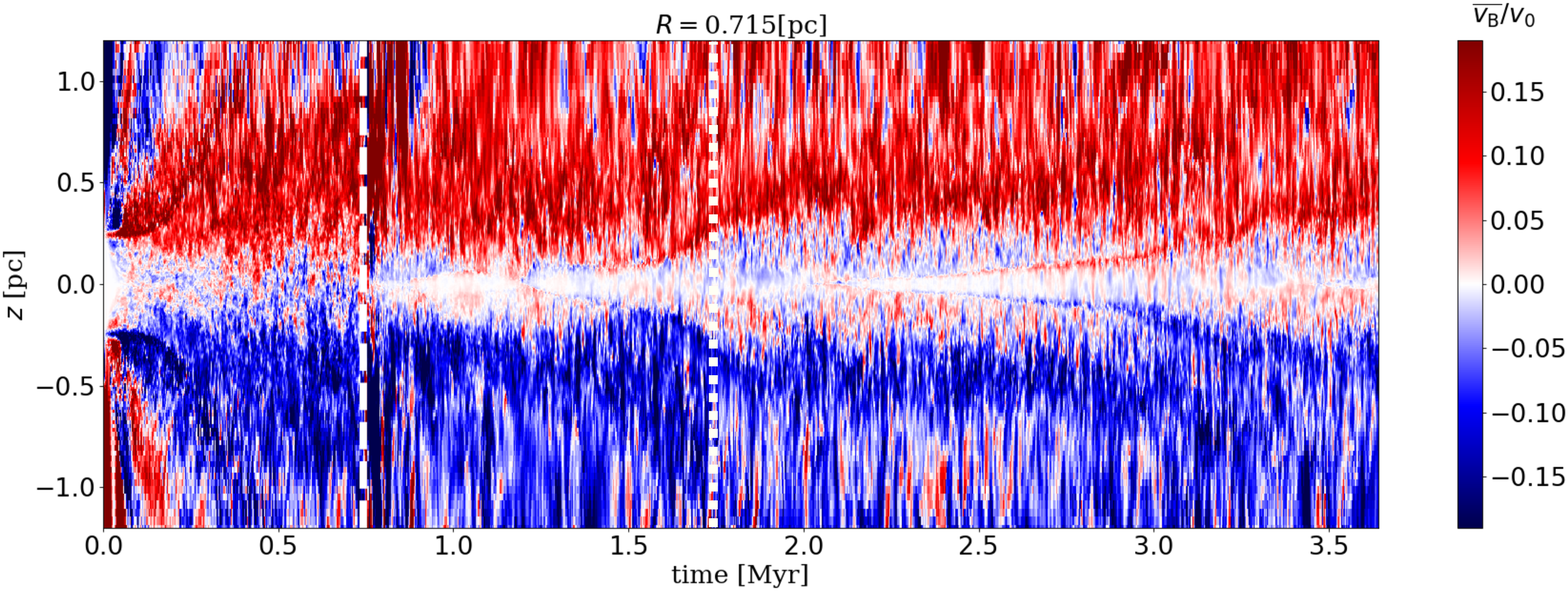}
\caption{
Same as Figure \ref{fig:512bfm}, but for the mean rise speed ($\overline{v_B}/v_0$) of the toroidal magnetic flux at $R=0.715$ pc.
Red and blue denote the positive speed and negative speed for the vertical direction, respectively.
\label{fig:pfav}}
\end{figure*}

The vertical escape of the mean field is described as the vertical magnetic energy transport using the Poynting flux $F_{P,~z}$,
\begin{equation}
F_{P,~z} = B_R \left( v_z B_{R} -  B_z v_R \right) - B_{\varphi} \left( v_{\varphi} B_z - B_{\varphi} v_z \right),
\end{equation}
\begin{equation}
\overline{v_B} \equiv \frac{ \overline{F_{P,~z}} }{ \overline{B}_{\varphi}^2 }, \label{eq:vB}
\end{equation}
where a rise speed $v_B$ denotes the vertical derivative of the pattern of the mean toroidal field \citep{2016MNRAS.457..857S}.
The large rise speeds occur at the direction reversal, and large fall speeds appear around the midplane and $|z|>0.5$ pc (Figure \ref{fig:pfav}).
The maximum rise speed $v_B$ is 37.3 km s$^{-1}$ in $|z|<0.8$ pc.
In the same way, Alfv\'{e}n speed and vertical speed are measured as $|\bm{v}_A| \sim 41.6$ km s$^{-1}$ and $|v_z|\sim 44.0$ km s$^{-1}$, and these speeds are comparable.
The rise speed cannot reach the gravitational escape speed, $244.8$ km s$^{-1}$ at $R=0.715$ pc.

\subsection{Turbulent Velocity Field and Magnetoconvective Instability}  \label{subsec:33}
\begin{figure}[h!]
\epsscale{0.8}
\plotone{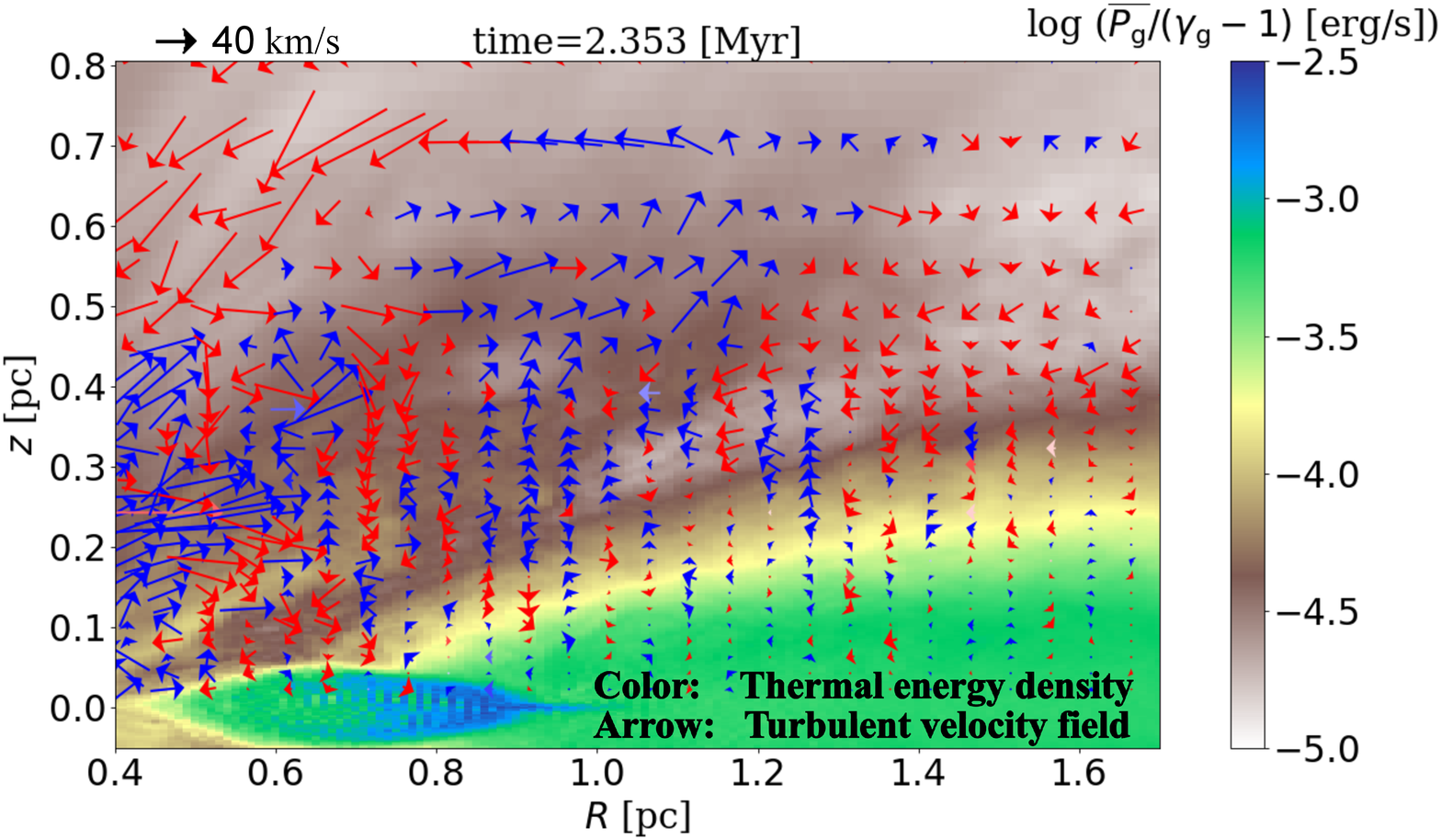}
\caption{
Turbulent velocity field (i.e. deviation from the mean velocity field) and mean thermal energy density distributions in the $R$-$z$ plane at the same time as the right panels of Figure \ref{fig:te}. The arrows above $z=0$ are colored blue for $\delta v_{z}>0$, and red for $\delta v_{z}<0$. (For reference, a velocity of 40 km s$^{-1}$ is shown on the top left.)
}
\label{fig:dist_Rz}
\end{figure}

In this subsection, we investigate the physical origin of the turbulent magnetic field shown in Section \ref{subsec:31} and how it is maintained.
Figure \ref{fig:dist_Rz} shows the turbulent velocity field of the gas at $t = 2.353$ Myr.
The maximum upflow (blue arrows) is about 18$\%$ of the escape velocity at that position.
The gas circulates with downward and upward flows in the ambient, where $\log \left\{ P_{\rm g}/(\gamma_{\rm g} -1) \right\} \lesssim -4$.
The velocity inside the disk is relatively smaller than that in the ambient.
The magnitude of the turbulent velocity for $R \lesssim 0.9$ pc is comparable to the sound speed.
That for $R \gtrsim 0.9$ pc is roughly 10$\%$ of the sound speed.
The direction of vectors does not coincide with the turbulent and mean magnetic fields; however, part of the gas falls into the midplane along the magnetic field lines.
Each component of the turbulent field and velocity has a difference only within one order of magnitude.

\begin{figure}[h!]
\epsscale{0.4}
\plotone{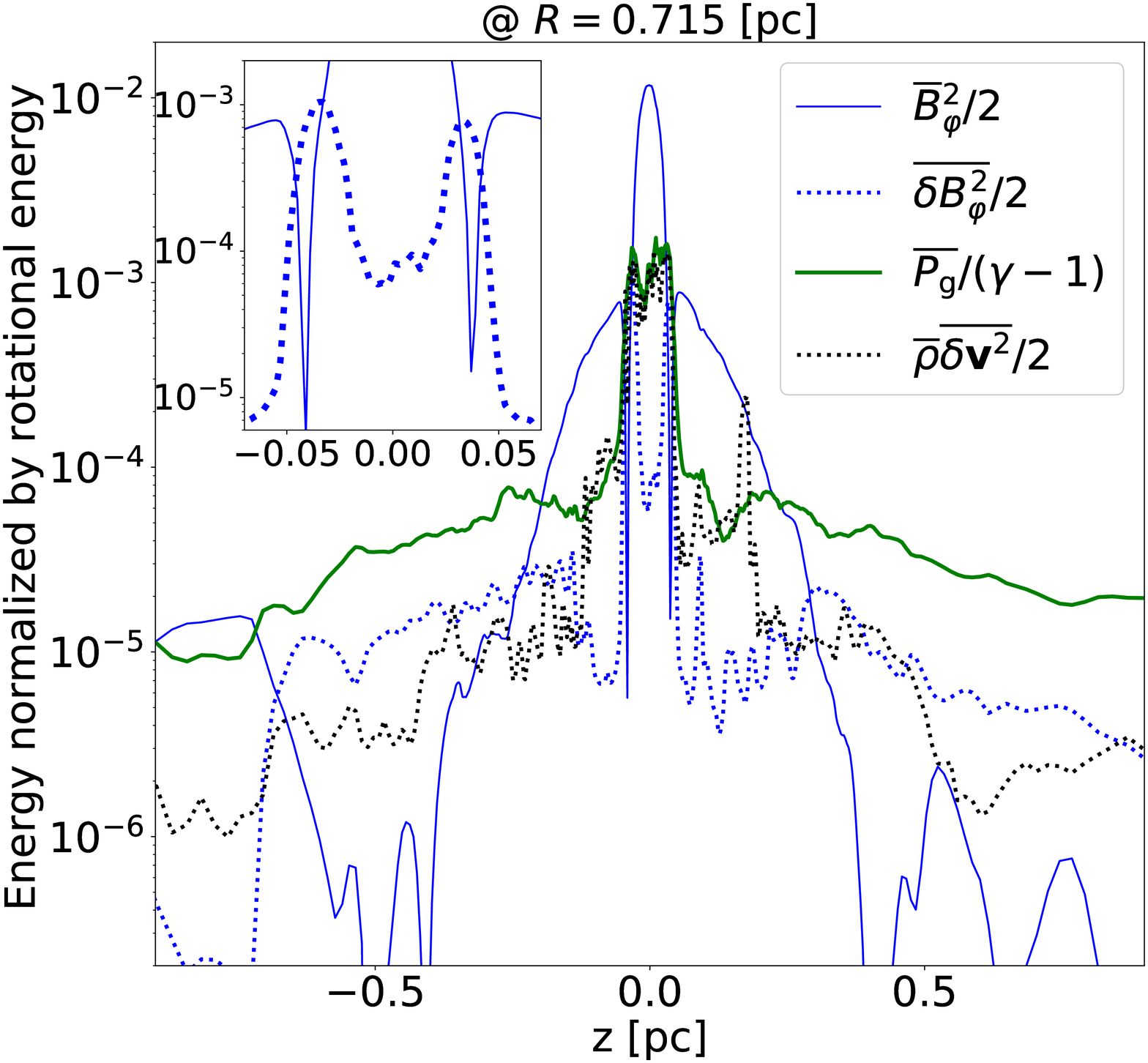}
\caption{Vertical distribution of mean energy densities in MHD+cooling/heating state at $t=2.353$ Myr. The mean energy density in the turbulent fields $\delta B_i^2$ and $\rho \delta v_i^2$ ($i=R, \varphi, z$) is averaged over the azimuthal direction. Normalized energy densities are the rotational energy, $\rho_0 v_K/2$, estimated by substituting the number density $n=10^2$ cm$^{-3}$ and Keplerian rotation $v_K$ at $R=0.715$ pc.
Colors are the thermal energy (green), the turbulent kinetic energy (black), and the toroidal magnetic energy (blue), respectively.
Dotted and solid lines denote the turbulent and mean components.
\label{fig:distz0790}}
\end{figure}

Figure \ref{fig:distz0790} shows the vertical structures of the magnetic, thermal, and kinetic energies at a given radius at $t= 2.533$ Myr.
Around the midplane, the mean toroidal magnetic field dominates the turbulent magnetic and kinetic energies ($\overline{\rho} \overline{|\delta \bm{v}|^2}/2$),
both of which are comparable to the thermal energy ($\overline{P_{\rm g}}/(\gamma_{\rm g} -1)$).
The mean field decreases more rapidly than the turbulent component with $z$.
As a result, there are regions where $\overline{\delta B_{\varphi}^2} > \overline{B}_{\varphi}^2$ ($0.2 \lesssim |z| $ pc $\lesssim 0.8$).
The transition between the turbulent component and the mean field also occurs at the disk surface ($|z| \sim 0.05$ pc), where
the mean field reverses its direction as seen in Figure \ref{fig:512bfm}.
At higher latitude ($|z| \gtrsim 0.3$ pc),
contrary to the midplane ($z = 0$ pc), turbulent toroidal field energy ($\overline{\delta B_{\varphi}^2}/2$) is about 10 $\%$ of the mean thermal energy,
but it is comparable to or a few times larger than the kinetic energy.

The vertical random motion could be related to the magnetoconvective instability
(or interchange instability, see, e.g., \citealt{Acheson1979}).
The unstable criterion is
\begin{eqnarray}
\frac{d}{dz} \left( \frac{P_{B}}{\rho^{\gamma_{B}}} \right) =\frac{d}{dz} \left( \frac{B_{\varphi}^2/2}{\rho^{2}} \right) <0 ~~~ ({\rm unstable}).  \label{eq:interchange}
\end{eqnarray}
This criterion corresponds to the convective instability for the gas
with decreasing specific entropy $s$ with large $|z|$, i.e.
\begin{eqnarray}
\frac{ds }{dz} \propto \frac{d}{dz} \left( \frac{P_{\rm g}}{\rho^{\gamma_{\rm g}} } \right) < 0 ~~~ ({\rm unstable}).   \label{eq:convection}
\end{eqnarray}
Replacing $P_{\rm g}$ and $\gamma_{\rm g}$ in the relation (\ref{eq:convection}) with $P_{B}$ and $\gamma_{B}$,
where the magnetic pressure $P_{B}=|\bm{B}|^2/2$, the criterion (\ref{eq:interchange}) is obtained.
Here $\gamma_{B}=1+P_B/U_B = 2$ for the energy density $U_B=|\bm{B}|^2/2$ \citep[e.g.,][]{Kulsrud2005}.
The criterion (\ref{eq:interchange}) can be written as $d\left( |B_{\varphi}|/\rho \right) /dz <0$. Therefore, the instability occurs when the mass frozen in the magnetic flux tube per unit length increases with increasing $|z|$.

\begin{figure}[h!]
\epsscale{0.45}
\plotone{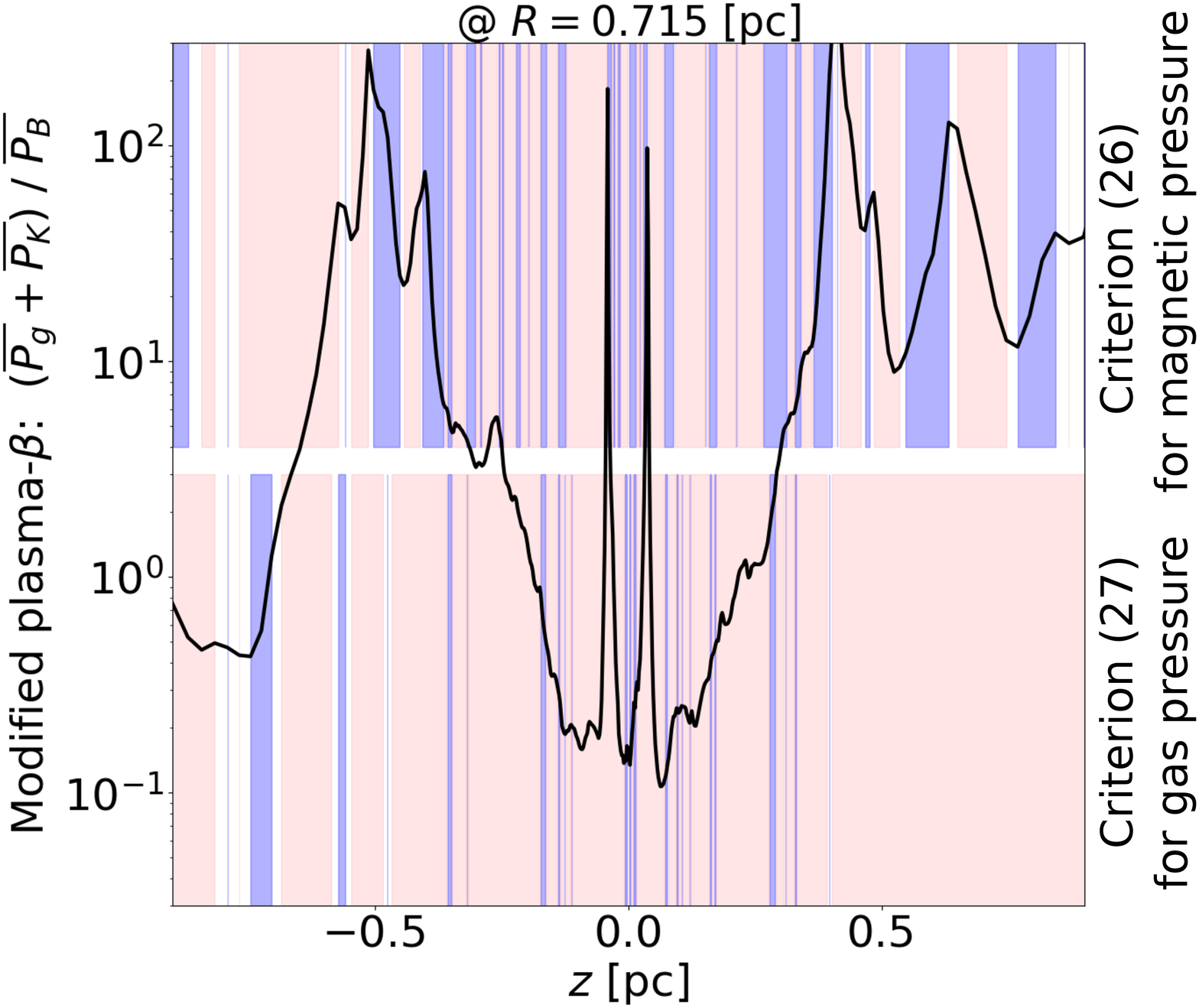}
\caption{Vertical distribution of the mean modified plasma-$\beta$ (solid line) and the criterion of convective instabilities (shadows) at $t=2.353$ Myr.
The mean modified plasma-$\beta$ represents $\left( \overline{P_{\rm g}} + \overline{P_{\rm K}} \right)/ \overline{P_{B}}$, where $P_{K}$ is the ram pressure evaluated by the turbulent kinetic energy, $ \rho |\delta {\bm{v}}|^2 $.
The vertical distribution stability is shaded in red, and the unstability in blue.
Stability evaluation is performed using the gradient of the mean quantities from Equation \ref{eq:interchange} (upper) and Equation \ref{eq:convection} (lower).
\label{fig:convection}
 }
\end{figure}

One should note that the criteria (\ref{eq:convection}) and (\ref{eq:interchange}) are necessary conditions for convection.
We compare the two criteria with the vertical variation of the modified plasma
$\hat{\beta}$, i.e. $\hat{\beta} \equiv \left( \overline{P_{\rm g}} + \overline{P_{\rm K}} \right)/ \overline{P_{B}}$ in Figure \ref{fig:convection},
where ${P_{\rm K}}\equiv \overline{\rho} \delta {\bm{v}}^2$.
Note that the strong ram pressure exists around the midplane, and $P_{\rm K} \sim P_{\rm g}$, but magnetic energy dominates.
In the bottom panel of Figure \ref{fig:convection}, hydrodynamic convection is essentially stable, i.e. the positive entropy gradient shown in shadows in red shadows.
In contrast, as seen in the top panel of Figure \ref{fig:convection}, there are many unstable regions for
the magnetoconvective instability (blue shadows).
From the conditions, i.e.
 $\frac{d}{dz} \left( \frac{P_B}{\rho^2} \right) <0$(unstable), $\frac{d}{dz} \left( \frac{P_{\rm g}}{\rho^{\gamma_{\rm g}}} \right) >0$ (stable),
 after some algebraic calculations to vanish the density gradient, we can derive
\begin{eqnarray}
\frac{1}{P_{B}} \frac{d P_{\rm g}}{dz} + \frac{\gamma_{\rm g}}{2} P_{\rm g} \frac{d}{dz} \left( \frac{1}{P_{B}} \right) >0.
\end{eqnarray}
When $d |B_{\varphi}|/dz \sim 0$ (i.e. when field direction reversal occurs) and/or $\gamma_{\rm g} /2 \sim 1$, this condition satisfies $d \beta /dz > 0$.
Figure \ref{fig:convection} shows that the regions with $d \hat{\beta} /dz > 0$ correspond to unstable regions for magnetoconvective instability.

\begin{figure}[h!]
\epsscale{0.5}
\plotone{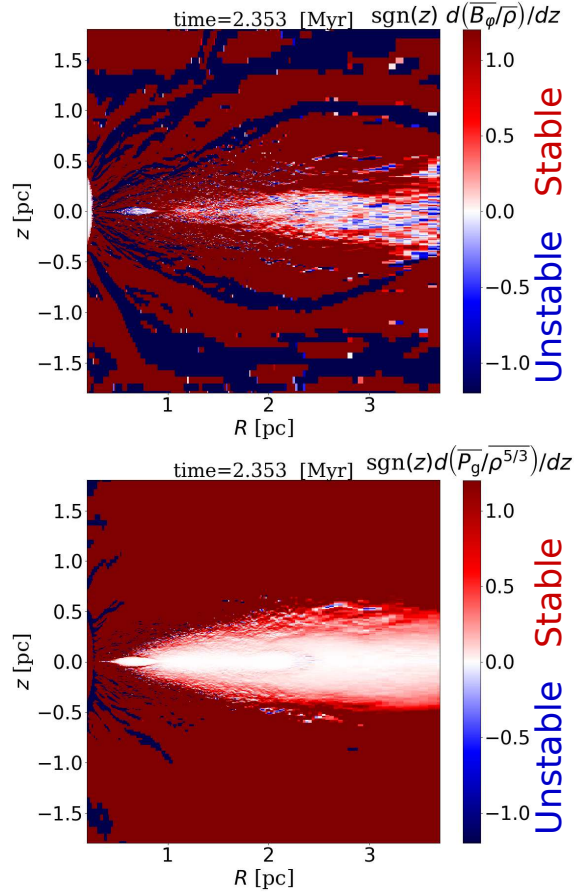}
\caption{Stability of convective instabilities in the $R$-$z$ plane at $t=2.353$ Myr.
Top: magnetoconvection of Equation \ref{eq:interchange}. Bottom: hydrodynamic convection of Equation \ref{eq:convection}.
Blue and red denote the same criteria as in Figure \ref{fig:convection}.
\label{fig:unstable}
}
\end{figure}

The two panels of Figure \ref{fig:unstable} show the unstable regions for the two criteria, (\ref{eq:interchange}) and (\ref{eq:convection}).
They show that the unstable regions for the magnetoconvection form
belt-like layers, where the mean toroidal field direction reverses (see, $B_{\varphi}$ on the $R$-$z$ plane in Figure \ref{fig:mag_0790}).
The unstable layers move with the rising direction pattern of the mean toroidal field.
The width of the unstable layers remains approximately constant around $0.05$-$0.1$ pc.
The bottom panel implies that the system is convectively stable.

\subsection{Thermal state of the magnetic activity}  \label{subsec:34}

\begin{figure}[h]
\epsscale{0.4}
\plotone{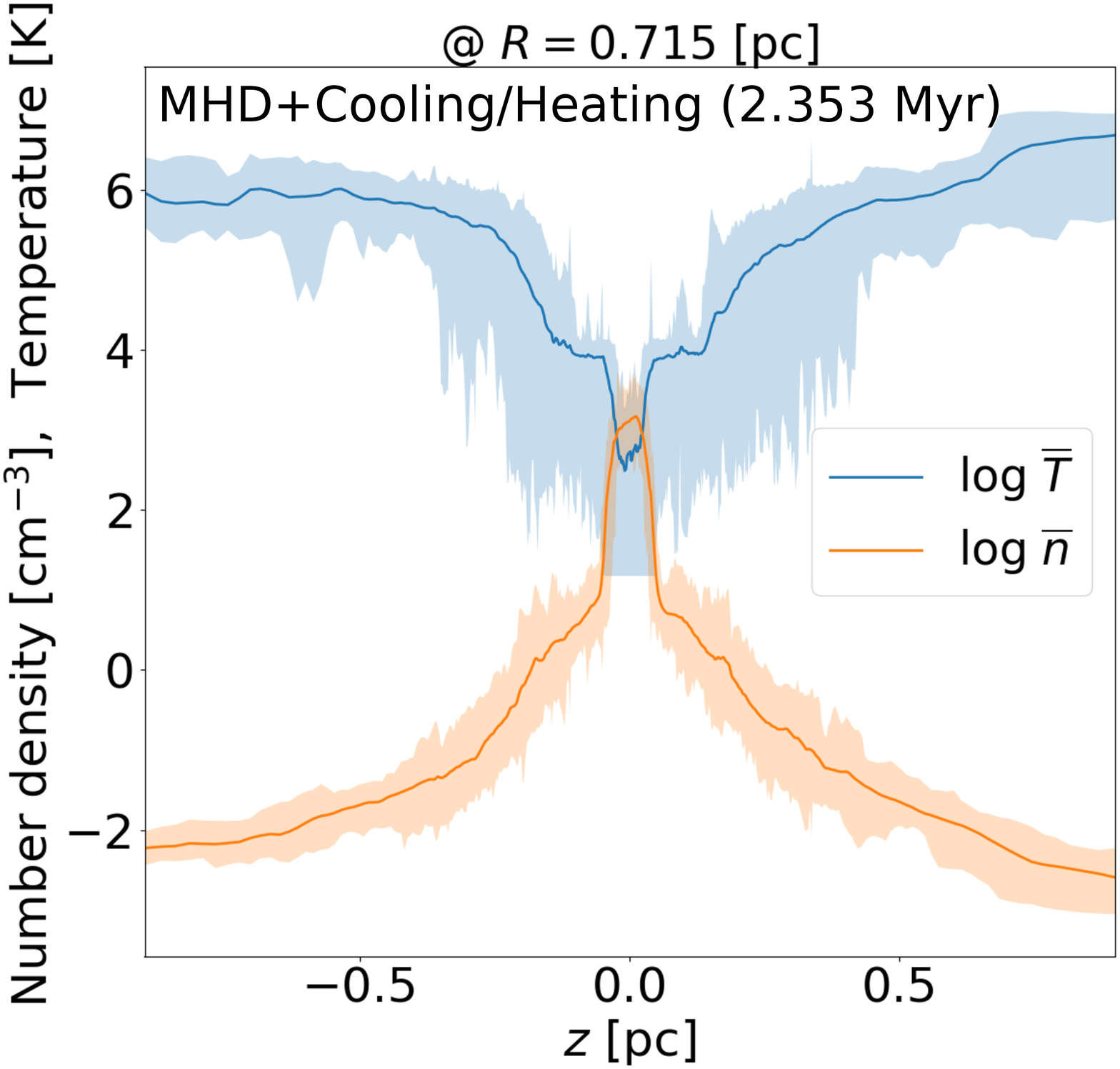}
\caption{Vertical distribution of the temperature (blue) and the number density (red) at $R=0.715$ pc of the MHD+cooling/heating state $t=2.353$ Myr. Shadowed areas are the maximum and minimum values over the toroidal direction. Solid lines are the azimuthally averaged density and temperature.
\label{fig:distztero}}
\end{figure}

Figure \ref{fig:distztero} shows the vertical distributions of the density and temperature at $R = 0.715 $ pc at $t = 2.353$ Myr.
For $|z| < 0.05 $ pc, the gas is cold ($\sim 8 \times 10^2$ K) and dense ($n \sim 10^3$ cm$^{-3}$).
As seen in Figure \ref{fig:te}(c), this cold disk extends to $R \sim 2 $ pc.
Outside this cold, dense disk, the gas is warm ($\sim10^4$ K) and less dense ($n \sim 10$ cm$^{-3}$);
therefore, they are roughly in pressure equilibrium.
The temperature of the ambient region ($|z| > 0.25$ pc) increases continuously until $\sim 10^6$ K.
Figure \ref{fig:distztero} also shows that there is a large azimuthal fluctuation around the mean values.
The density fluctuation is in the order of unity or less while the temperature, as a maximum, fluctuates by three orders of magnitude.
For the cold disk ($|z|<0.05$ pc), the minimum temperature reaches the lower limit, and the maximum does not exceed $10^4$ K of the warm disk.

\begin{figure}[h!]
\centering
\epsscale{0.4}
\plotone{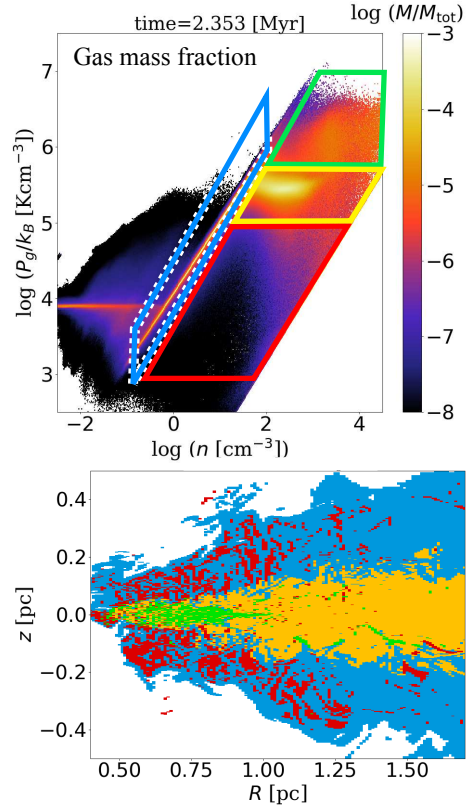}
\caption{Top panel is a 2D histogram with the gas pressure in the vertical axis and the number density in the vertical axis at $t=2.352$ Myr. Color contour denotes the mass fraction distribution occupying cells of $\Delta (\log n) = \Delta (\log P_{\rm g}/k_B) =0.01$ and averaged over the number of mesh points in each $P_{\rm g}$-$n$ space cell. Bottom panel is the $R$-$z$ plane distribution corresponding to some regions enclosing each color in the top panel. Light blue: $3.4<\log (T$[K])$< 4.6$, $-1<\log (n$[cm$^{-3}$]$)< 2$, red: $2.0 < \log T < 3.4$, $3<\log(P_{\rm g}/k_B$[Kcm$^{-3}$]$)< 5$, yellow: $1 < \log T < 3.4$, $5<\log(P_{\rm g}/k_B)< 5.8$, and green: $1 < \log T < 3.6$, $5.8<\log(P_{\rm g}/k_B)< 6.8$. Note that the gas temperature denotes a constant slope as $d \log (P_{\rm g}/k_B)/d \log n$, assuming the ideal gas.
\label{fig:ph2sp} }
\end{figure}

To quantify the temperature and density fluctuations in phase space, we plot the gas mass fraction
as a function of gas pressure and number density in the top panel of Figure \ref{fig:ph2sp}.
A large amount of gas mass is collected in characteristic regions over the gas mass fraction $\log (M/M_{\rm tot}) > -5 $.
These are multiphase states created by the thermal instability.
The gas is also distributed in a wide range of densities and temperatures, but a large fraction of the gas is in a state with $P_{\rm g}/k_B \gtrsim 10^5$ and $n \gtrsim 1$.
The bottom panel in Figure \ref{fig:ph2sp} shows the spatial distributions of four thermal states shown in the top panel on the $R$-$z$ plane.
The colors represent the region enclosed by the same colors on $P_{\rm g}$-$n$ plane.
The low-temperature gases is patchy in the warm ($\sim10^4$ K) thick disk shown in blue.
The mass fraction of this low-temperature gas is small.
Although the mass fraction of gas in the red region is small,
The green region refers to the dense gas in a cold ($\sim100$ K) phase.
The gas shown in yellow in the top panel mostly forms the thick disk at $R > 0.9$ pc as shown in the bottom panel, where they are approximately isothermal (therefore, the gas pressure is constant).

\begin{figure}[h!]
\epsscale{1.2}
\plotone{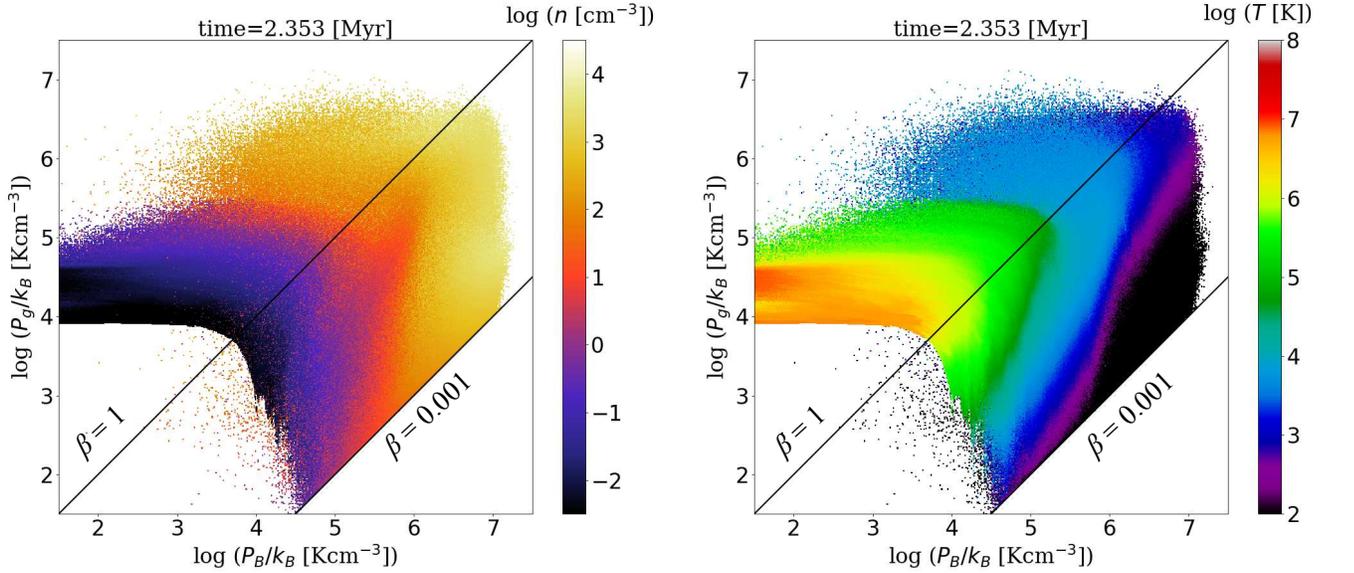}
\caption{2D histograms of the gas pressure in the vertical axis and the magnetic pressure in the horizontal axis inside $R=0.9$ pc at $t=2.352$ Myr. Left: number density, Right: temperature. Occupying cells are $ \Delta (\log P_{\rm g}/k_B) = \Delta (\log P_{B}/k_B) = 0.01$. Solid lines are $\beta=1.000$ and $0.001$ (lower limit).
\label{fig:pb_tero}}
\end{figure}

In Figure \ref{fig:pb_tero}, density and temperature for $R < 0.9$ pc (i.e., the quasi-steady cold, thin gas disk) are plotted as a functions of the magnetic pressure $P_{\rm B}$ and the thermal pressure $P_{\rm g}$ at $t = 2.35$ Myr.
In the region where magnetic pressure dominates ($\beta < 1$),
the number density of the left panel is distributed in a wide range from $n \sim 10^{-2}$ to $10^4$ cm$^{-3}$ for a given gas pressure.
It shows that the temperature is lower than $T \sim 10^5$ K, and
the strongly magnetized gas consists of cold gas with $T < 1000$ K, which dominates the total mass (Figure \ref{fig:ph2sp}).
In the high-$\beta$ domain, the temperature and density fixed at gas pressure are not sensitive to changes in magnetic field.
The thermal state is determined by the compression and expansion of the gas pressure.

In the above results, we observe a thin cold ($ T < 100$ K) disk where the magnetic field is strong ($\beta  \lesssim 1$) at $R < 0.9$ pc (see Figures \ref{fig:ph2sp} and \ref{fig:pb_tero}).
This structure is stable at least until the end of the simulation, i.e. $ t \sim 3.6$ Myr ( $\sim 96$ rotational periods at $R = 1$ pc).
The disk is maintained by the MRI and by the mass inflow from the outer disk ($R > 0.9$ pc), where the gas is less dense and less magnetized.
Therefore, we suspect that this strongly magnetized disk is not “transient”, and it could last during the lifetime of  the AGNs ($\sim 10$ Myr).

If this is the case, a strong magnetic field at $R < 1$ pc could be expected in AGNs, and
it can be observable by the Zeeman effect using future observations by ALMA,
ng-VLA, and SKA.

If the mass supply from the outer region is stopped, the magnetic field can be dissipated by the ambipolar diffusion (see, Equation \ref{eq:amb} and the Appendix \ref{appendix}). The expected time scale is
\begin{equation}
t_{\rm Adiff} \sim Re_{\rm M, A} ~ \frac{L}{V} \sim 3.8 ~ {\rm Myr}  ~ \left( \frac{x}{10^{-4}} \right) \left( \frac{B}{1 {\rm~ mG}} \right)^{-2}  \left( \frac{n_n}{10^4 {\rm ~cm}^{-3}} \right)^{2} \left( \frac{L}{1 {\rm~ pc}} \right)^2,
\end{equation}
which is still long enough to be observed.

\section{Discussion} \label{sec:4}

\subsection{Direction reversal in the low-$\beta$ MRI} \label{subsec:41}

We found that the low-$\beta$ disk formed by radiative cooling and heating is discernible by
the direction reversal (Fig. \ref{fig:512bfm}) and vertical transfer of the magnetic field (Fig. \ref{fig:pfav}).
A long reversal period is observed in the strong toroidal field with low-$\beta$ compared to the turbulence with high-$\beta$.
This trend is also found in the local 3D simulations of an isothermal gas \citep{2013ApJ...767...30B, 2016MNRAS.457..857S}.
The mean plasma-$\beta$, $\overline{\beta} \sim 0.1-0.4$ in the midplane is also similar to our result \citep[see, Fig. \ref{fig:convection}; Table 2 of][]{2016MNRAS.457..857S}.
Compared to an analytical model \citep{2015ApJ...809..118B}, it was demonstrated that heating efficiency, defined as the ratio between the dissipation rate
(e.g. the magnetic reconnection) and mean toroidal field production rate, decreases with $\beta$.
\citet{2016MNRAS.457..857S} showed that assuming steady Poynting flux,
the period of the direction reversal is proportional to the rotational period ($T_{\rm cycle}=2\pi /\Omega \xi_B^{-1} $)
and is determined by the phenomenological parameters; $\xi_B=\nu \eta_B /(\beta - \nu +1)$,
where $v_B = \eta_B \Omega z$ (see also Equation \ref{eq:vB}) and $\nu$ is the degree of turbulent heating.
The turbulent heating caused by the nonlinear MRI is weakened by the small plasma-$\beta$ \citep[see, Table 3 of][]{2016MNRAS.457..857S}.

\begin{figure*}[h!]
\epsscale{1.2}
\plotone{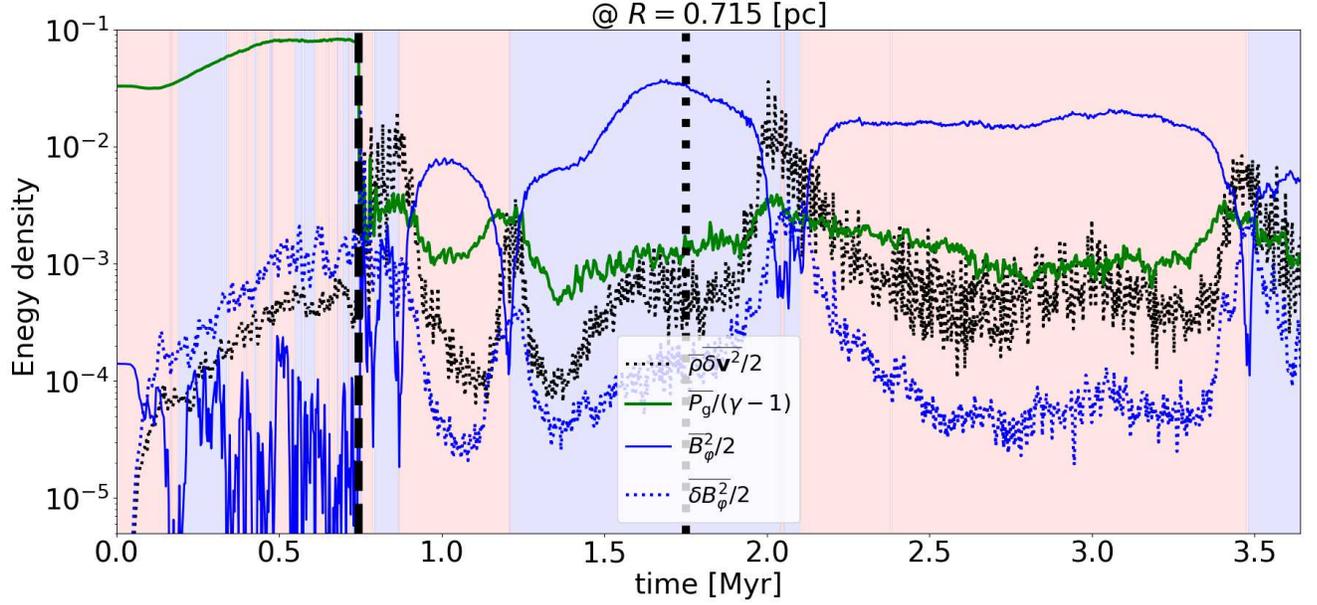}
\caption{Time variation of the energy densities in the midplane at $R=0.715$ pc.
Each energy density is normalized by the rotational energy with a density of $\rho_0$ and Keplerian rotation at a radius of $R=0.715$ pc.
Solid and dotted lines represent the energy density of the mean and turbulent components, respectively.
Different colors of solid lines denote the mean toroidal B-field energy (blue), the mean thermal energy (green), and the turbulent kinetic energy (gray).
Shaded colors of red and blue signify the direction of the mean toroidal field.
\label{fig:ene}}
\end{figure*}

\begin{figure*}[h!]
\epsscale{1.15}
\plotone{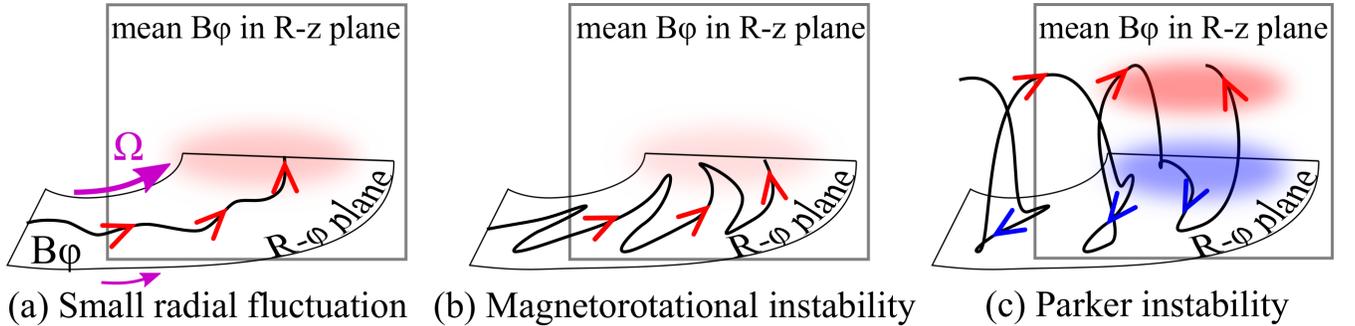}
\caption{Schematic illustration of the MRI-Parker dynamo in the differential rotational disk.
A magnetic field line is drawn as black curves with colored arrows. Red and blue arrows signify parallel and antiparallel and the purple arrows indicate the rotational direction.
(a) A magnetic field line is threaded by the toroidal field with a small radial turbulence on the $R$-$\varphi$ plane at a certain height;
(b) MRI and disk rotation stretch the magnetic field line to the radial and toroidal direction in $R$-$\varphi$ plane;
(c) Parker instability buoyantly escapes from the $R$-$\varphi$ plane of (a).
\label{fig:dynamo}}
\end{figure*}

The field direction reversal as seen in Figure \ref{fig:512bfm} is caused by the changes in balance between the turbulent energy and the mean field energy.
When the direction of the mean field, which is represented by two shaded colors, reverses (e.g., $t=1.20, 2.12, 3.48$ Myr in Figure \ref{fig:ene}),
the energy density $\overline{B}_{\varphi}^2/2$
(blue solid line) temporarily drops and the turbulent magnetic field ($\delta \bm{B}_{\varphi}^2/2$; blue dotted line) is amplified.
While the turbulent kinetic energy ($\overline{\rho} \delta \bm{v}^2/2$; black dotted line) exceeds the thermal energy ($\sim P_{\rm g}$; green solid line), the turbulent magnetic energy (blue dotted line) is smaller
than the thermal energy of cold gas ($T\sim1000$ K), i.e. $\beta$ at this moment becomes large.
The direction reversal is caused by a combination of MRI and the Parker instability \citep[Figure \ref{fig:dynamo} or see, e.g.,][]{2013ApJ...764...81M}.
A magnetic field line with a small radial fluctuation (panel (a)) is stretched to the radial and azimuthal directions (panel (b)).
The stretched field line with a longer wavelength is selectively buoyed up toward higher latitudes by the Parker instability (panel (c)).
As a result, for the $R$-$\varphi$ plane, the buoyant field line has an opposite direction to the field line in the midplane.
In the low-$\beta$ disk, direction reversal is possible by driving turbulence from the mean field; thus, the process to high $\beta$ will be important (i.e. magnetoconvection in Section \ref{subsec:33}).
Direction reversal was widely observed in adiabatic MHD simulations \citep[e.g., ][]{Beckwith2011, 2011ApJ...736..107O, 2012ApJ...744..144F, 2013ApJ...764...81M, 2013ApJ...763...99P, 2016ApJ...826...40H}, but we found here for the first time that {\it direction reversal also occurs in the low-$\beta$ disk with the multiphase gas} ($10<T$ [K]$<10^5$).

We observed the strongly magnetized disk driven by MRI in Section 3.
We discuss here what constrains the strength of the magnetic field (plasma-$\beta$) in this simulation.
In Figure \ref{fig:convection},  the mean plasma-$\beta$ around the cold disk shows $\bar{\beta} > 0.1$.
This satisfies the condition for the MRI, i.e. $\bar{\beta} > \beta_{\rm crit}$ \citep{Begelman2007}, above which the MRI can be driven, and
\begin{eqnarray}
\beta_{\rm crit} \sim 0.01 \left( \frac{T}{1000  {\rm~ K}}  \right)^{1/2}   \left( \frac{v_{\rm K}}{253 {\rm~ km s}^{-1}} \right)^{-1}.  \label{eq:MRIcrit}
\end{eqnarray}
where we adopted the Keplerian rotation $v_K$ at $R=0.715$ pc and the mean temperature in Figure \ref{fig:distztero}.
On the other hand, the mean plasma-beta is expressed as $\bar{\beta}^{-1} \sim (\delta B_{\varphi}^2/2 + \overline{B_{\varphi}}^2/2)/P_{\rm g}$.
For $\overline{B_{\varphi}}^2  \geq \delta B_{\varphi}^2 $, which is the case in the disk, $\bar{\beta} \leq P_{\rm g}/ \delta B_{\varphi}^2 $.
From Figure 15, $P_{\rm g} / \delta B_{\varphi}^2  \sim 10$, therefore, the mean plasma-beta can be constrained as $0.01 \lesssim \bar{\beta}  \lesssim 10$.

It has been known that the MRI-driven turbulence depends on numerical spatial resolution.
\citet{2011ApJ...738...84H,2013ApJ...772..102H} applied the quality factor $Q_{\varphi} \equiv \lambda/(R\Delta\varphi)$,
which is the resolution of the characteristic MRI wavelength ($\lambda=2 \pi B_{\varphi}/\sqrt{\rho\Omega^2}$).
They identified the empirical condition as $Q_{\varphi} \ga 20 $.
However, for the toroidal field, the numerical convergence of adiabatic turbulence has not been clarified yet.
The low-$\beta$ MRI in cold gas requires high resolution,
\begin{eqnarray}
 Q_{\varphi} \sim 32  \left( \frac{N_{\varphi}}{512}  \right) \left( \frac{v_{\varphi}}{207 {\rm~ km \, s^{-1}}} \right)^{-1} \left( \frac{T}{100 {\rm~ K}} \right)^{1/2}  \left( \frac{\beta}{10^{-2}} \right)^{-1/2}, \label{eq:Qvalue}
\end{eqnarray}
where $N_{\varphi}=2\pi/\Delta \varphi$ is the azimuthal resolution.
In our model, $N_{\varphi}=512$ and we found that $Q_{\varphi} \ga 20$ \citep[see, e.g.][]{Kudoh2018}.
Additionally, a comparison with $N_{\varphi}=128$ showed that the low resolution is $Q_{\varphi}<10$, and thus, there is no direction reversal in the cold and low-$\beta$ disk.
Our simulations are in agreement with the estimation of Equation \ref{eq:Qvalue} and the empirical condition.
In the long-term calculation, the turbulent magnetic field is sensitive to the numerical flux solver and the high-order accuracy.
In order to reduce numerical dissipation, HLLD flux solver is employed \citep[see, e.g.][]{2013ApJ...772..102H}.
\citet[][e.g., Section 3.2.1; 4.3]{2019PASJ...71...83M} pointed out that the high-order accuracy prevents the dissipation rather than the low-order scheme.
We took a highly precise numerical approach, hence the differences from previous global 3D simulations.

\subsection{Radiation pressure} \label{subsec:42}

AGNs emit enormous energy fluxes over a wide wavelength, and its feedback is important for
 the dynamics of the circumnuclear gas \citep[e.g.,][]{2016ApJ...825...67C,Chan2017}.
However, IR radiative pressure is not dynamically effective \citep{2016MNRAS.460..980N}, and cannot contribute to MRI in the cold gas.
Notably, the anisotropic radiation pressure on the dust in the gas with
the X-ray heating produces flows in a fountain-like manner
in the central tens-of-parsec pc regions around the AGNs  \citep{Wada2012c}.
The radiation pressure is expected to be stronger than the magnetic pressure, i.e.
\begin{eqnarray}
\frac{ P_{\rm rad} }{ P_{B} } &\sim& \frac{mn L_X \kappa_d \alpha_d / \left( c R \right)}{B_{\varphi}^2/2} \\ \nonumber
&\sim& 3 \left( \frac{L_X}{10^{-4} L_{\rm Edd}} \right)    \left( \frac{\kappa_d \alpha_d}{10^3 {\rm~ cm}^2 {\rm g}^{-1}} \right)
\left( \frac{n}{10^{3} {\rm~ g cm}^{-3}} \right)
\left( \frac{R}{1 {}\rm~ pc} \right)^{-1}  \left( \frac{B_{\varphi}}{1 {\rm~ mG}} \right)^{-2},  \label{eq:rOm}
\end{eqnarray}
where $\kappa_d=10^5$ is the dust opacity and $\alpha_d=0.01$ is the gas to dust ratio \citep{Wada2012c}.
In a subsequent paper, we will investigate the effect of radiation pressure, how magnetic structures are
changed, and how magnetic buoyancy or magnetoconvection can help.

\section{Summary} \label{sec:5}

We studied the evolution of a magnetized multiphase gas using global 3D MHD simulations in the parsec-scale galactic nuclei.
The simulation starts from an adiabatic state (${\cal L}=0$ in Equation \ref{eq:energy})
with a weak toroidal field, $\beta=100$, until the MHD turbulence is fully developed ($\beta \sim 0.6$) for $\sim $ 25 rotational periods at $R=1$ pc.
Thereafter, the effects of the radiative cooling and the X-ray heating from the accretion disk around the SMBH are taken into account for an additional 97 rotational periods (2.89 Myr).
The magnetic pressure dominated disk is formed due to MRI cooperating with the radiative cooling.
The quasi-steady state in a time sufficiently longer than the dynamical timescale is attained in the radii $R<0.9$ pc for $t>1.75$ Myr, as confirmed by a time variation of the mean toroidal field and a constant of the accretion rate averaged over variation timescale.

Major findings are:

(i)
The cold ($<10^3$ K) gas forms a geometrically thin disk around the midplane. The warm ($\sim 10^4$ K) gas forms a
thicker disk, and the hot ($\sim 10^{6}$ K) gas is distributed to higher latitudes (Figures \ref{fig:distztero}, \ref{fig:ph2sp}).
The mass fraction ($>10^{-5}$) on the $P_{\rm g}$-$n$ plane is in the warm and cold phases (Figure \ref{fig:ph2sp}).
The magnetic pressure is stronger (i.e. $\beta (<1)$) in the cold, dense gas (Figure \ref{fig:pb_tero}).

(ii)
The mean magnetic field is dominated by a toroidal component, and a strongly magnetized cold disk
with $\beta \sim 0.02$ is formed (Figures \ref{fig:te}, \ref{fig:mag_0250}, and \ref{fig:mag_0790}).
The mean toroidal field moves with the cold gas to a radial and vertical direction.
The energy of the turbulent field is suppressed by the cooling effect; however, it is always comparable to or smaller than the thermal energy (Figures \ref{fig:512Emag} and \ref{fig:ene}).

(iii)
The turbulent motion in the multiphase gas is observed in the $R$-$z$ plane (Figure \ref{fig:dist_Rz}).
Magnetoconvective instability plays a key role in maintaining turbulence for a long period.
The unstable condition $d(|\overline{B}_{\varphi}|/\rho)/dz<0$ (Equation \ref{eq:interchange}) coincides with the region in the modified plasma beta increasing vertically upward (Figure \ref{fig:distz0790}) and in the belt-like mean field reversal in the $R$-$z$ plane (Figure \ref {fig:unstable}).
The transition between the turbulent component and the mean field also occurs at the disk surface ($|z| \sim 0.05$ pc and $0.2 \lesssim |z|$ pc$\lesssim 0.8$), where
the mean field spatially reverses its direction (Figures \ref{fig:mag_0790} and \ref{fig:distz0790}).

(iv)
The quasi-steady state differs for the plasma $\beta$ inside the disk.
The high-$\beta$ disk is achieved by the saturation of $|\delta \bm{B}|^2$ and the oscillation of $\overline{B}_{\varphi}^2$ (Figure \ref{fig:512Emag}).
In our simulations starting from the initial weak toroidal field ($\beta=100$), the magnetic field strength amplified by the MRI remains $\beta\ga 1$ and the oscillation timescale is about 10 rotational periods, as mentioned in the previous studies.
The quasi-steady state in the low-$\beta$ disk is obtained by the periodic change of $\overline{B}_{\varphi}^2$.
The period in the low-$\beta$ state is more than 5 times longer (about 50 rotational periods) than that found in the high-$\beta$ state.

(v)
We found that even for low $\beta$ $(\sim 0.02)$, the mean toroidal field shows direction reversals with time (Figures \ref{fig:512Emag} and \ref{fig:512bfm}).
This is caused by the transportation of the magnetic field vertically due to Parker instability (Figure \ref{fig:dynamo}), similar to the adiabatic state.
The direction reversal of the mean $B_\varphi$ occurs, when the turbulent magnetic energy becomes larger than mean magnetic energy (Figure \ref{fig:ene}).
The moving speed of the magnetic field (Equation \ref{eq:vB}) estimated by the Poynting vertical flux is about 10$\%$ of the rotation speed.
This speed becomes a maximum where the mean toroidal field direction reverses (Figure \ref{fig:pfav}).
To continue this cycle, mean magnetic flux transport from midplane to vertical direction is important.

\acknowledgments
The authors are grateful to the anonymous referee for constructive comments and suggestions.
We thank Ryoji Matsumoto, Mami Machida, and Yusuke Tsukamoto for their constructive comments and discussion, and the CANS+ developer team for the numerical techniques.
Numerical computations were performed on Cray XC50 and XC30 systems at the Center for Computational Astrophysics, National Astronomical Observatory of Japan.
This work was supported by JSPS KAKENHI grant No. 16H03959 and by NAOJ ALMA Scientific Research grant No. 2020-14A.

 \section*{Additional Links}
Movies of snapshots of Figures 2 and 4 are available in the following link \\
\url{https://astrophysics.jp/MHD_torus/}.

\appendix

\section{Nonideal MHD effects}  \label{appendix}

 In this paper, we solved ideal MHD equations, whereby the gas in the circumnuclear region moves together with
 the magnetic field. This would be justified because we assumed that
 the ionization degree is $(x\equiv n_i/n_n)$, $ x \sim 10^{-4} $ following \cite{Meijerink2005}.
 However, this assumption is incorrect if the magnetic Reynolds numbers for the dissipation processes
 in terms of the ohmic effect, Hall effect, and ambipolar diffusion being smaller than unity.
Here, we confirm this.

The electric field in the induction equation (Equation \ref{eq:induction}) is replaced by the generalized Ohm's law \citep[e.g.,][]{Braginskii1965},
\begin{eqnarray}
\bm{E} + \bm{v} \times \bm{B}
= \eta \bm{J} + \eta_{\rm H} \left( \bm{J} \times \frac{\bm{B}}{|\bm{B}|} \right)+ \eta_{\rm A} \bm{J}_{\perp},  \label{eq:generalOhm}
\end{eqnarray}
where $\bm{J}_{\perp}$ is the current perpendicular to the magnetic field.
The three terms in the r.h.s of Equation \ref{eq:generalOhm} are the magnetic dissipation of the ohmic, Hall, and ambipolar terms, respectively.
The coefficients are formulated as follows:
\begin{eqnarray}
\eta=\frac{m_e \left( \nu_{\rm ei} + \nu_{\rm en} \right)}{4 \pi  e^2 n_e},
\end{eqnarray}
\begin{eqnarray}
\eta_{\rm H}=\frac{ |\bm{B}| }{4 \pi e n_e},
\end{eqnarray}
\begin{eqnarray}
\eta_{\rm A}=\left( \frac{\rho_{\rm n}}{\rho}  \right)^2 \frac{ |\bm{B}|^2 }{4 \pi \left( \rho_{\rm i} \nu_{\rm in} + \rho_{\rm e} \nu_{\rm en} \right)},
\end{eqnarray}
where indices $i, e, n$ denote the particle species of ion, electron, and neutral hydrogen, respectively.
Here, $\nu_{a b}$ denotes the collisional frequency of a particle "a" with a particle "b".
Collisional frequencies are given by \cite{Spitzer1962} assuming elastic collision,
\begin{eqnarray}
\nu_{\rm in}= 5 \times 10^{15} x^{-1} n_e \sqrt{\frac{8 k_{\rm B} T}{\pi} \frac{m_i+m_n}{m_i m_n} }, \\
\nu_{\rm en}=  10^{15} x^{-1} n_e \sqrt{\frac{8 k_{\rm B} T}{\pi} \frac{m_e+m_n}{m_e m_n} },\\
\nu_{\rm ei}=  \frac{(4 \pi)^2 e^4 \ln \Lambda}{3 m_e^2} n_e \left[ \frac{ m_e }{2 \pi k_B T}  \right]^{ \frac{3}{2}},
\end{eqnarray}
where the Coulomb logarithm $\ln \Lambda$ is about one order of magnitude.
The electron-neutron collision is not effective,
$\nu_{en}/\nu_{ei} \sim 0.52/ \ln \Lambda (10^4/x)  (T/10^3 {\rm ~K})^2 < 1$.
In dense gas, $\nu_{in}= 3.5 \times 10^{13} \rho_n$ is often used \citep{Draine1983}.

The magnetic Reynolds number ($Re_{\rm M} \equiv LV/ \eta$) is defined as
the ratio of the $|\bm{v} \times \bm{B}|$ term to the dissipation term in Equation \ref{eq:generalOhm}.
We adopt the typical advection scale, $L \sim 1$ pc and $V \sim v_{\rm A} \sim 10$ km s$^{-1}$,
and the magnetic Reynolds numbers in each dissipation are,
 \begin{eqnarray}
Re_{\rm M} &\sim& 5.43 \times 10^{14}  \left( \frac{x}{10^{-4}} \right)^{3/2} \left( \frac{\beta}{0.01} \right)^{-1/2}  \left( \frac{B}{1 ~{\rm mG}} \right)^{-1} \left( \frac{n_n}{10^4 ~{\rm cm}^{-3}} \right)^{1/2} \left( \frac{V}{10 {\rm~ km \, s^{-1}}} \right) \left( \frac{L}{1 {\rm~ pc}} \right),  \label{eq:Ohm}
\end{eqnarray}
\begin{eqnarray}
Re_{\rm M, H} \sim 8.5 \times 10^8 \left( \frac{x}{10^{-4}} \right)  \left( \frac{B}{1 ~{\rm mG}} \right)^{-1} \left( \frac{n_n}{10^4 {\rm~cm}^{-3}} \right) \left( \frac{V}{10 {\rm~km\, s^{-1}}} \right) \left( \frac{L}{1 {\rm~pc}} \right), \label{eq:Hall}
\end{eqnarray}
\begin{eqnarray}
Re_{\rm M, A} \sim 37.9  \left( \frac{x}{10^{-4}} \right) \left( \frac{B}{1 {\rm~ mG}} \right)^{-2}  \left( \frac{n_n}{10^4 {\rm ~cm}^{-3}} \right)^{2}  \left( \frac{V}{10 {\rm~ km \, s^{-1}}} \right) \left( \frac{L}{1 {\rm~ pc}} \right).  \label{eq:amb}
\end{eqnarray}
Equations \ref{eq:Ohm} and \ref{eq:Hall} imply that we can ignore
the ohmic and Hall dissipation.
The ambipolar diffusion may be important for very strong magnetic fields (e.g. $\gg 1$ mG) and/or diffuse media ($n_n < 100$ cm$^{-3}$),
excepted for high ionization, e.g. $x \ga 10^{-4}$.

\bibliography{AGNMRI}{}
\bibliographystyle{aasjournal}



\end{document}